\documentclass[journal]{IEEEtran}
\usepackage{amsmath,amsfonts,amsthm,mathtools}
\usepackage{algorithmic}
\usepackage{array}
\usepackage[caption=false,font=normalsize,labelfont=sf,textfont=sf]{subfig}
\usepackage{textcomp}
\usepackage{stfloats}
\usepackage{url}
\usepackage{verbatim}
\usepackage{graphicx}
\usepackage{xcolor}
\usepackage{nomencl}
\usepackage{etoolbox}
\usepackage{ifthen}
\usepackage{booktabs}
\usepackage{adjustbox}
\makenomenclature
\usepackage{cite}
\usepackage{multirow}
\usepackage{bm}
\usepackage{makecell}
\usepackage{enumitem}

\allowdisplaybreaks 
\usepackage[subtle,tracking=normal]{savetrees}

\def\BibTeX{{\rm B\kern-.05em{\sc i\kern-.025em b}\kern-.08em
    T\kern-.1667em\lower.7ex\hbox{E}\kern-.125emX}}
\usepackage{balance}
\newtheorem{theorem}{Theorem}

\newtheorem{remark}{Remark}

\usepackage{algorithm}

\newcommand{\rev}[1]{\textcolor{blue}{#1}} 
\usepackage{url}
\usepackage{hyperref} 
\usepackage{xcolor}

\hypersetup{
    colorlinks=true, 
    linkcolor=blue, 
    filecolor=black, 
    urlcolor=blue, 
    citecolor=blue, 
    linkbordercolor={1 1 1}, 
    pdfborder={0 0 0} 
}

\begin{document}

\bstctlcite{IEEEexample:BSTcontrol}
\title{Grid-Aware Real-Time Dispatch of Microgrid\\ with Generalized Energy Storage: A Prediction-Free Online Optimization Approach}

\author{\IEEEauthorblockN{Kaidi Huang, \textit{Graduate Student Member, IEEE}}, \IEEEauthorblockN{Lin Cheng, \textit{Senior Member, IEEE}},
\IEEEauthorblockN{Ning Qi, \textit{Member, IEEE}},\\
\IEEEauthorblockN{David Wenzhong Gao, \textit{Fellow, IEEE}},
\IEEEauthorblockN{Asad Mujeeb, \textit{Graduate Student Member, IEEE}},
 \IEEEauthorblockN{Qinglai Guo, \textit{Fellow, IEEE}}

 \thanks{Manuscript created xxx, 2024; revised xxxx, 2025; revised xxxx, 2025; accepted xxxx 2025.
 This work was supported in part by the National Key Research and Development Program of China (No. 2021YFE0191000), National Natural Science Foundation of China (No. 52037006) and China Postdoctoral Science Foundation special funded project (No. 2023TQ0169). Paper no.TSG-01682-2024. (\textit{Corresponding author}: Ning Qi.)

 Kaidi Huang, Lin Cheng, Asad Mujeeb, and Qinglai Guo are with the Department of Electrical Engineering, Tsinghua University, Beijing 100084, China (e-mail: hjd23@mails.tsinghua.edu.cn).
 
 Ning Qi is with the Department of Earth and Environmental Engineering, Columbia University, New York, NY
10027 USA (e-mail: nq2176@columbia.edu).
 
  David Wenzhong Gao is with the Department of Electrical and Computer Science, University of Denver, Colorado, USA (e-mail: Wenzhong.2001.Gao@ieee.org).}
}

\markboth{IEEE TRANSACTIONS ON Smart Grid,~Vol.~X, No.~X, XX August~2025}
{How to Use the IEEEtran \LaTeX \ Templates}

\maketitle

\begin{abstract}
    This paper proposes a novel prediction-free two-stage coordinated dispatch framework for the real-time dispatch of grid-connected microgrid with generalized energy storages (GES). The proposed framework explicitly addresses grid awareness, non-anticipativity constraints, and the time-coupling characteristics of GES, providing microgrid operators with a near-optimal, reliable, and adaptable dispatch tool. In the offline stage, we generate the hindsight state-of-charge (SoC) trajectories of GES by solving the multi-period economic dispatch with historical scenarios. Subsequently, leveraging this historical information (SoC trajectories, net loads, and electricity prices), we synthesize and dynamically update online references for both SoC and opportunity cost through kernel regression. We propose an \rev{adaptive Lagrange multiplier-based} online convex optimization algorithm, which innovatively incorporates reference tracking for global vision and expert-tracking for step-size updates. We provide theoretical proof to show that the proposed OCO algorithm achieves a sublinear bound of both dynamic regret and time-varying hard constraint violation. Numerical studies using ground-truth data from the Australian Energy Market Operator demonstrate that the proposed method outperforms state-of-the-art methods, reducing operational costs by 5.0–6.2\% and voltage violations by 0.8–9.1\%. These improvements mainly result from mitigating myopia by reference tracking and the adaptive capability provided by dynamically updated references and \rev{adaptive Lagrange multipliers}. \rev{Sensitivity analysis demonstrates the robustness, computational efficiency, and scalability of the proposed method.}
\end{abstract}


\mbox{}
\renewcommand{\nomgroup}[1]{%
	\item[\textbf{%
		\ifthenelse{\equal{#1}{A}}{\textit{A.\ Abbreviations}}{}%
		\ifthenelse{\equal{#1}{B}}{\textit{B.\ Indices and Sets}}{}%
		\ifthenelse{\equal{#1}{C}}{\textit{C.\ Parameters}}{}%
		\ifthenelse{\equal{#1}{D}}{\textit{D.\ Decision Variables}}{}%
  	\ifthenelse{\equal{#1}{E}}{\textit{E.\ Functions}}{}%
	}]%
}
\nomenclature[A]{SoC}{State of charge}
\nomenclature[A]{OCO}{Online convex optimization}
\nomenclature[A]{GES}{Generalized energy storage}
\nomenclature[A]{MPC}{Model predictive control}
\nomenclature[A]{DP}{Dynamic Programming}

\nomenclature[B]{$\mathcal{D}/\mathcal{S}$}{$\text{Sets for diesel generator units}/$GES units}
\nomenclature[B]{$\mathcal{B}/\mathcal{T}$}{$\text{Sets for buses}/$time periods}
\nomenclature[B]{$m\text{,}n/t$}{$\text{Indices for buses}/$time periods}
\nomenclature[B]{$d/s$}{$\text{Indices for diesel generator units}/$GES units}
\nomenclature[B]{$i$}{Indices for experts in the OCO algorithm}
\nomenclature[B]{$u$}{Indices for historical scenarios}

\nomenclature[C]{$\alpha_{i,t}/\beta_{i,t}$}{Step sizes for the expert $i$ in the OCO algorithm}
\nomenclature[C]{$\epsilon$}{Probabilistic level of chance-constraints}
\nomenclature[C]{$\varepsilon_{s}$/$\eta_{s}$}{$\text{Self-discharge rate}/$efficiency of GES unit $s$ at time $t$}
\nomenclature[C]{$P_{l\text{,}t}$/$Q_{l\text{,}t}$}{$\text{Active}/$reactive power of load $l$ at time $t$}
\nomenclature[C]{$c_{s}^{+}$/$c_{s}^{-}$}{$\text{Discharge}/$charge cost of  of GES unit $s$}
\nomenclature[C]{$c_d$/$c_{g\text{,}t}$}{$\text{Fuel cost of diesel generator unit $d$}/$electricity price at time $t$}
\nomenclature[C]{$R_{nm}$/$X_{nm}$}{$\text{Resistance}/$reactance of line $nm$}
\nomenclature[C]{$Z_{nm}$}{Impedance of line $nm$}
\nomenclature[C]{$\overline{V}_{m}$/$\underline{V}_{m}$}{$\text{Upper}/$lower bounds of squared voltage of bus $m$}
\nomenclature[C]{$\overline{P}_d$/$\underline{P}_d$}{$\text{Upper}/$lower power bound of diesel generator unit $d$}
\nomenclature[C]{$\overline{P}_{s\text{,}t}^{+}$/$\overline{P}_{s\text{,}t}^{+}$}{$\text{Upper discharge}/$charge power bound of GES unit $s$ at time $t$}
\nomenclature[C]{$\overline{P}_{g}$}{Upper bound of power imported from the grid}
\nomenclature[C]{$\overline{E}_{s\text{,}t}$/$\underline{E}_{s\text{,}t}$}{$\text{Upper}/$lower SoC bound of of GES unit $s$ at time $t$}
\nomenclature[C]{$\varsigma_{s\text{,}t}$}{Baseline consumption of GES unit $s$ at time $t$}
\nomenclature[C]{$\varphi$}{Penalty coefficient for reference tracking}
\nomenclature[C]{$\nu_{i\text{,}t}$}{\rev{Adaptive Lagrange multiplier} for the expert $i$ at time $t$ in the OCO algorithm}
\nomenclature[C]{$\mu/\sigma$}{$\text{Mean}/$standard deviation of the distribution}
\nomenclature[C]{$\rho_{i\text{,}t}$}{Weight of the expert $i$ in the OCO algorithm}

\nomenclature[C]{$\ell_{i\text{,}t}$}{Surrogate loss of the expert $i$ at time $t$}

\nomenclature[C]{$\hat{E}_{s\text{,}t}/\hat{\lambda}_{t}$}{Reference for SoC trajectory of GES unit $s$ at time $t$$/$opportunity cost at time $t$}

\nomenclature[C]{$P_{l\text{,}t}$/$Q_{l\text{,}t}$}{$\text{Active}/$reactive power of load $l$ at time $t$}
\nomenclature[C]{$\tau$}{Bandwidth of kernel regression}

\nomenclature[C]{$N/M$}{$\text{Number of experts}/$scenarios}

\nomenclature[D]{$P_{s\text{,}t}^{+}$/$P_{s\text{,}t}^{-}$}{$\text{Active discharge}/$charge power of GES unit $s$ at time $t$}
\nomenclature[D]{$Q_{s\text{,}t}^{+}$/$Q_{s\text{,}t}^{-}$}{$\text{Reactive discharge}/$charge power of GES unit $s$ at time $t$}
\nomenclature[D]{$E_{s\text{,}t}$}{SoC of GES unit $s$ at time $t$}
\nomenclature[D]{$P_{d\text{,}t}$/$Q_{d\text{,}t}$}{$\text{Active}/$reactive power of diesel generator unit $d$ at time $t$}
\nomenclature[D]{$P_{g\text{,}t}$}{Active power imported from the grid at time $t$}
\nomenclature[D]{$p_{m\text{,}t}$/$q_{m\text{,}t}$}{$\text{Active}/$reactive power injection at bus $m$, time $t$}
\nomenclature[D]{$P_{nm\text{,}t}$/$Q_{nm\text{,}t}$}{$\text{Active}/$reactive power flow of line $nm$ at time $t$}
\nomenclature[D]{$I_{nm\text{,}t}$/$V_{m\text{,}t}$}{$\text{Squared current of line $nm$}/$squared voltage of bus $m$ at time $t$}

\nomenclature[E]{$\mathbb{P}$}{Probability function}
\nomenclature[E]{$F^{-1}$}{Inversed cumulative distribution function}
\nomenclature[E]{$f_{t}$}{Objective function}
\nomenclature[E]{$h_{t}$}{Constraint function}
\nomenclature[E]{$K_{t}$}{Kernel function}
\nomenclature[E]{$\left\langle\text{ } \right\rangle$ }{Standard inner produc function}

\printnomenclature

\section{Introduction}\label{Introduction}

\IEEEPARstart{M}{icrogrid} enables the integration and coordination of various resources, including renewables, energy storages, diesel generators, controllable loads, power imported from the main grid, etc. It is critical to develop efficient dispatch methods to fully leverage these resources, ensuring reliable local demand supply while maintaining grid security. However, significant challenges remain, primarily due to highly stochastic uncertainties arising from renewables, loads, and market prices, as well as the inherent time-coupling characteristics introduced by energy storage.

Early studies primarily focus on day-ahead dispatch and rely on two-stage optimization methods, such as (two-stage) robust optimization~\cite{tan2024robust}, stochastic optimization~\cite{antoniadou2022scenario}, chance-constrained optimization~\cite{qi2023chance}, and distributionally robust optimization~\cite{qiu2023two}. However, two-stage optimization methods assume that the second-stage (real-time) decisions are made after all the uncertainties are observed, which contradicts the non-anticipativity~\cite{feng2023day} that the real-time decisions can only be determined based on uncertainty information available up to the current time, without any knowledge of future realizations. Moreover, the day-ahead dispatch becomes conservative and lacks adaptability to the highly volatile real-time environment. This limitation motivates the exploration of online optimization methods to real-time dispatch by leveraging continuously updated uncertainty information obtained from either forecasts or real-time observations.

The first class of online optimization methods is model predictive control (MPC)~\cite{hu2023economic,yang2023tracking,guo2023long}, which repeatedly solves a finite-horizon optimization problem using real-time forecasts in a rolling-horizon manner. Although MPC is an industrially mature method, it suffers from a myopic perspective due to its limited prediction horizon and heavy reliance on forecast accuracy, which has been evidenced in~\cite{guo2023long} that myopic decisions and inaccurate forecasts result in severe feasibility issues in the long run. The second class of online optimization methods is dynamic programming (DP), which leverages Bellman's principle of optimality to decompose the multi-period optimization into multiple single-period optimizations via the value function. The value function is trained offline based on historical data, and the online decision-making is made based on the current uncertainty observation. Among various DP variants, stochastic dual dynamic programming (SDDP) has been most widely applied due to its effectiveness in handling high-dimensional uncertainties~\cite{aaslid2022stochastic,bhattacharya2016managing,lan2022fast}. However, SDDP still faces challenges in value function construction and the ``curse of dimensionality'', limiting its applicability to microgrid dispatch with multiple energy storages. To mitigate these limitations, approximate dynamic programming (ADP)~\cite{li2021multi,shuai2018stochastic} and reinforcement learning (RL)~\cite{hu2022soft,stai2022computing} have emerged, offering computationally efficient solutions through approximate value functions or learning-based policies. However, ADP methods heavily depend on the choice of approximation methods and value function initialization, and RL methods suffer from unstable training processes and online performance.

In addition to the aforementioned methods, there is growing interest in employing online algorithms from the control domain, notably Lyapunov optimization and online convex optimization (OCO)~\cite{wang2023online}, as they are prediction-free and provide theoretical performance guarantees (e.g., regret bounds), offering distinct advantages over MPC and DP methods. Lyapunov optimization~\cite{shi2015real,stai2020online,liu2017online} adopts a ``1-lookahead" decision pattern~\cite{wang2023online}, where decisions are made sequentially by observing current uncertainties and minimizing a drift-plus-penalty term derived from the Lyapunov function to balance immediate costs and queue stability. However, the online policy derived from Lyapunov optimization may become myopic if uncertainties vary significantly over time, as it utilizes neither forecasts nor historical data. Moreover, the ``observe-then-act" pattern is unsuitable for practical microgrid dispatch, since decisions typically need to be made before uncertainties are realized. For instance, when the microgrid participates in the real-time market, energy trading decisions with the main grid must be made prior to market clearing, when electricity prices become known. And dispatch decisions must be made at the beginning of each interval, when the non-stationary uncertainties remain unpredictable or unknown. In contrast, OCO is specifically designed for the ``0-lookahead" decision-making pattern,  making proactive decisions before uncertainties are revealed, and has recently gained increasing attention in power systems, particularly in applications such as demand-side management and ancillary services~\cite{kim2016online,zhao2020distributed,lesage2019online}. However, the inherent limitation of OCO in microgrid dispatch lies in:

\begin{enumerate}
    \item \textbf{Myopia and limitations in handling time-coupling constraints.} Most existing OCO algorithms~\cite{yu2020low,yi2021regret,cao2020constrained,ding2021dynamic,muthirayan2022online,yi2022regret,kim2016online,zhao2020distributed,lesage2019online,guo2022online} leverage only the information from the previous period for online decision-making, leading to myopic decisions and focusing primarily on single-period constraints, thus failing to handle the time-coupling constraints inherent in energy storage. Our previous work~\cite{qi2025long} introduces kernel regression to learn the long-term state-of-charge (SoC) trajectory, and employs penalty terms within the proposed OCO algorithm to track this trajectory. Although tracking the SoC trajectory can mitigate myopic behavior, it may not be effective for short-duration energy storage (e.g., battery). This is because it does not accurately capture the truthful opportunity cost of battery, which depends strongly on both the SoC and volatile electricity prices~\cite{xu2020operational}. In contrast, off-grid hydrogen storage exhibits relatively stable future opportunity costs. Therefore, it is preferable to explicitly learn and track both the SoC trajectory and the opportunity cost in real-time dispatch.
    \item \textbf{Simplified assumptions and grid unawareness.} Many existing OCO algorithms rely on simplified problem settings that are unsuitable for realistic microgrid dispatch. OCO algorithms address two performance metrics: \textit{Regret} and \textit{Constraint Violation} and aims to achieve sublinear bounds on both with respect to the time horizon $T$. Regret quantifies the cumulative difference between the realized objective incurred by the OCO algorithm and the hindsight optimum, and constraint violation measures the cumulative feasibility violations resulting from the ``0-lookahead" decision pattern. However, some studies~\cite{yu2020low,cao2020constrained,muthirayan2022online,yi2022regret} only evaluate \textit{static regret} and assume the fixed hindsight optimum over time, thus failing to address dynamic performance under the volatile market environment. Others either ignore constraint violations entirely~\cite{kim2016online,zhao2020distributed,lesage2019online,qi2025long} or consider only time-invariant constraints (fixed constraints over time) or soft constraint violations (net cumulative violation allowing feasible margins to offset)~\cite{ding2021dynamic,liu2022simultaneously}, potentially causing significant infeasibility issues and compromising grid security (e.g., limits on voltage and reverse power flow). Therefore, we should develop a novel OCO algorithm that explicitly addresses dynamic regret (with time-varying hindsight optimum) and time-varying hard (cumulative violation without accounting for feasible cases) constraint violations.
\end{enumerate}

\begin{table*}[!t]
    \centering
\setlength{\abovecaptionskip}{-0.1cm}
\setlength{\belowcaptionskip}{-0.1cm}
    \renewcommand\arraystretch{1.25} 
    \setlength{\tabcolsep}{0cm} 
    \caption{Comparison of this paper with other related literature}\label{comparison table}
    \begin{center}
    \begin{tabular}{cccccc}
    \toprule
    \multirow{2}[2]{*}{Reference} & \multirow{2}[2]{*}{Method} & \multirow{2}[2]{*}{Prediction-Free (Information)} & \multirow{2}[2]{*}{Myopic (Horizon)} & \multirow{2}[2]{*}{Regret (Bound)} & \multirow{2}[2]{*}{Constraints Violation (Bound)} \\
          &       &       &       &       &  \\
    \midrule
\cite{tan2024robust,antoniadou2022scenario,qi2023chance,qiu2023two}          & Two-Stage & $\times$ (Day-Ahead Forecast) & $\times$ (Day-Ahead) & ——    & —— \\
\cite{hu2023economic,yang2023tracking}          & MPC   & $\times$(Real-Time Forecast) & $\checkmark$ (Short Rolling) & ——    & —— \\
\cite{guo2023long}          & MPC   & $\times$ (SoC Reference+Real-Time Forecast) & $\times$ (Short Rolling) & ——    & —— \\
\cite{aaslid2022stochastic,bhattacharya2016managing,lan2022fast}          & SDDP  & $\times$ (Value Function+Probabilistic Forecast) & $\times$ (Single-Period) & ——    & —— \\
\cite{li2021multi,shuai2018stochastic,hu2022soft,stai2022computing}           & ADP/RL & $\times$ (Value Function+Real-Time Forecast) & $\times$ (Single-Period) & ——    & —— \\
\cite{shi2015real} & Lyapunov & $\checkmark$ (Drift-Plus-Penalty+1-Look-Ahead) & $\checkmark$ (Single-Period) &\rev{\makecell{Time-average gap ($O(1/V)$)\textsuperscript{1}}} & \rev{Virtual queue proxy ($O(V)$)\textsuperscript{1}} \\
\cite{stai2020online} & Lyapunov & $\checkmark$ (Drift-Plus-Penalty+1-Look-Ahead) & $\checkmark$ (Single-Period) &\rev{\makecell{$\text{UB}_{\text{iLypRC}}$\textsuperscript{2}}} & \rev{Explicit SoE bounds\textsuperscript{2}} \\
\cite{yu2020low}         & OCO   & $\checkmark$(0-Look-Ahead) & $\checkmark$ (Single-Period) &\makecell{Static ($O(\sqrt{T})$)}      &  \makecell{Time-Invariant+Soft ($O(1)$)}  \\
\cite{yi2021regret}          & OCO   & $\checkmark$ (0-Look-Ahead) & $\checkmark$ (Single-Period) &\makecell{Dynamic ($O(\sqrt{T(1+P_x)})$)}      &  \makecell{Time-Invariant+Hard ($O(\sqrt{T})$)}  \\
\cite{cao2020constrained}          & OCO   & $\checkmark$ (0-Look-Ahead) & $\checkmark$ (Single-Period) &\makecell{Static ($O(\sqrt{T})$)}      &  \makecell{Time-Varying+Soft ($O(T^{3/4})$)} 
\\
\cite{ding2021dynamic}          & OCO   & $\checkmark$(0-Look-Ahead) & $\checkmark$ (Single-Period) & \makecell{Dynamic ($O(\max(T^\delta P_x,T^{1-\delta}))$)}      &  \makecell{Time-Varying+Soft ($O(\max(T^{1-\delta},T^\delta))$)}
\\
\cite{muthirayan2022online}          & OCO   & $\checkmark$ (0-Look-Ahead) & $\checkmark$ (Single-Period) &\makecell{Static ($O(T^{\max\{1-a-c,c\}})$)}      &  \makecell{Time-Varying+Hard ($O(T^{1/2-c/2})$)}  \\
\cite{yi2022regret}          & OCO   & $\checkmark$ (0-Look-Ahead) & $\checkmark$ (Single-Period) &\makecell{Static ($O(T^{\max(\delta,1-\delta)})$)}      &  \makecell{Time-Varying+Hard ($O(T^{1-\delta/2})$)} 
\\
\cite{guo2022online}          & OCO   & $\checkmark$ (0-Look-Ahead) & $\checkmark$ (Single-Period) &\makecell{Dynamic ($P_x\sqrt{T}$)}      &  \makecell{Time-Varying+Hard ($O(T^{3/4})$)} 
\\
\cite{qi2025long}          & OCO   & $\checkmark$ (SoC Reference+0-Look-Ahead) & $\times$ (Single-Period) &\makecell{Dynamic ($O(\sqrt{T(1+P_x)})$)}      & ——  \\
This Paper & OCO   & $\checkmark$ (SoC$\&$OC Reference+0-Look-Ahead) & $\times$ (Single-Period) &\makecell{Dynamic ($O(T^{1/2+\chi}\sqrt{1+P_x})$)}      &  \makecell{Time-Varying+Hard ($O(\log_2(T)T^{1-\chi/2})$)} \\
    \bottomrule
    \end{tabular}
    \end{center}
$P_{x}$: path-length, i.e., the accumulated variation of optimal decisions; 
$P_{h}$: function variation, i.e., the accumulated variation of constraints; OC: opportunity cost. \rev{\textsuperscript{1}\textbf{Note on \cite{shi2015real}:} 
The regret is given as a time-average gap of $O(1/V)$ between the algorithm's cost and the infinite-horizon optimal cost. Constraint violation is not measured cumulatively but indirectly represented via the stability of virtual queues, which serve as proxies for long-term feasibility under time-coupled constraints. \textsuperscript{2}\textbf{Note on \cite{stai2020online}:} The regret is defined as the performance gap between the proposed algorithm and a perfect-foresight oracle, captured by an explicit upper bound $\text{UB}_{\text{iLypRC}}$. Constraint satisfaction is ensured by explicitly bounding the state-of-energy (SoE) through parameter-dependent feasibility guarantees. }

\end{table*}

We summarize the research gaps by comparing related literature with this paper in Table~\ref{comparison table}. To the best of our knowledge, no existing research has developed a prediction-free online optimization method for real-time microgrid dispatch that explicitly addresses grid awareness, non-anticipativity constraints, and time-coupling characteristics of energy storage, with rigorous proof on regret and constraint violation bounds.

Motivated by this background, this paper proposes a novel prediction-free two-stage coordinated dispatch framework for the real-time dispatch of grid-connected microgrid with generalized energy storages (GES). It provides microgrid operators with a near-optimal, reliable, and adaptable dispatch tool. Specially, our contributions are as follows:

\begin{enumerate}
    \item \textbf{Dispatch Framework:} We propose a novel prediction-free two-stage coordinated dispatch framework that effectively mitigates the myopia of online optimization and ensures robustness and strong adaptability without dependence on forecasts or hyperparameter tuning. In the offline stage, we generate hindsight SoC trajectories using historical scenarios. Subsequently, these SoC trajectories along with netloads and electricity prices are utilized to generate references for both SoC and opportunity cost of GES through kernel regression. These references are recursively updated online and explicitly integrated as tracking terms into the proposed OCO algorithm for real-time decision-making.
    \item \textbf{Solution Methodology:} We address the inherent uncertainties in GES using the chance-constrained optimization with a tractable deterministic reformulation. We propose an \rev{adaptive Lagrange multiplier-based} OCO algorithm, which innovatively incorporates reference tracking for global vision and expert-tracking for adaptive learning rate updates. Through designed algorithmic settings, we rigorously prove sublinear bounds for both dynamic regret and time-varying hard constraint violations.
    \item \textbf{Numerical Study:}  Numerical case studies based on ground-truth data from the Australian Energy Market Operator validate the effectiveness of the proposed method. Under identical reference tracking, our method outperforms state-of-the-art MPC and Lyapunov optimization methods, achieving operational cost reductions of 5.0\% and 6.2\%, and reducing voltage violations by 9.1\% and 0.8\%, respectively. Moreover, reference tracking substantially mitigates myopic nature of OCO, with the references for SoC and opportunity cost contributing approximately 6.5\% and 3.9\% to cost reductions, respectively. And dynamically updated references better capture real-time uncertainties compared to static day-ahead references, reducing operational costs by approximately 3.0\%. \rev{Sensitivity analysis demonstrates the robustness, computational efficiency, and scalability of the proposed method.}
\end{enumerate}

We organize the remainder of the paper as follows. Section~\ref{Two-stage} provides problem formulation and preliminaries of microgrid dispatch. Section~\ref{method} introduces the proposed prediction-free two-stage coordinated dispatch framework. Section~\ref{case study} presents numerical case studies based on ground-truth data. Finally, section~\ref{Conclusion}~concludes this paper.

\section{Problem Formulation and Preliminaries}\label{Two-stage}
\subsection{Oracle Multi-Period Economic Dispatch}
We consider a grid-connected microgrid that consists of renewables, diesel generators, energy storages, flexible loads, and fixed loads. Most flexible loads, such as thermostatically controlled loads and electric vehicles, exhibit energy storage characteristics, motivating the concept of virtual energy storage. To facilitate unified modeling, energy storage and virtual energy storage can be integrated into a GES framework. For a detailed transformation from physical flexible load models into the GES model, please refer to~\cite{qi2023chance}. The oracle mutli-period economic dispatch (OED) is formulated in~\eqref{OED}.
 \begin{subequations}\label{OED}
\begin{align}
   &\min \sum_{t\in\mathcal{T}}[\sum_{s\in\mathcal{S}}(c_{s}^{+}P_{s\text{,}t}^{+}+c_{s}^{-}P_{s\text{,}t}^{-})+\sum_{d\in\mathcal{D}}c_dP_{d\text{,}t}+c_{g\text{,}t}P_{g\text{,}t}]\label{objective}\\
    &\text{s.t. }\forall m\text{,}n\in\mathcal{B}\text{, }\forall d\in\mathcal{D}\text{, }\forall s\in\mathcal{S}\text{, }\forall t\in\mathcal{T}\nonumber\\
    & p_{m\text{,}t} = P_{nm\text{,}t}-R_{nm}I_{nm\text{,}t}-\sum_{w:m\rightarrow w}P_{mw\text{,}t}\label{P_flow}\\
    & q_{m\text{,}t} = Q_{nm\text{,}t}-X_{nm}I_{nm\text{,}t}-\sum_{w:m\rightarrow w}Q_{mw\text{,}t}\label{Q_flow}\\
    & p_{m\text{,}t} =\sum_{s\in\mathcal{B}_{m}}(P_{s\text{,}t}^{-}-P_{s\text{,}t}^{+})+\sum_{l\in\mathcal{B}_{m}}P_{l\text{,}t}-\sum_{d\in\mathcal{B}_{m}}P_{d\text{,}t}\label{P_input}\\
    & q_{m\text{,}t} =\sum_{s\in\mathcal{B}_{m}}(Q_{s\text{,}t}^{-}-Q_{s\text{,}t}^{+})+\sum_{l\in\mathcal{B}_{m}}Q_{l\text{,}t}-\sum_{d\in\mathcal{B}_{m}}Q_{d\text{,}t}\label{Q_input}\\
    &V_{m\text{,}t} = V_{n\text{,}t} - 2(R_{nm}P_{nm\text{,}t} + X_{nm}Q_{nm\text{,}t})+Z_{nm}^2I_{nm\text{,}t}\label{V_def}\\
    &\left\|(2P_{nm\text{,}t}\text{,}2Q_{nm\text{,}t}\text{,}V_{n\text{,}t} - I_{nm\text{,}t})^\top\right\|_2 \leq V_{n\text{,}t} + I_{nm\text{,}t}\label{SOCP}\\
    &\underline{V}_{m}\leq V_{m\text{,}t}\leq\overline{V}_{m}\label{V_limit}\\
    &0\leq I_{nm\text{,}t}\leq\overline{I}_{nm}\label{I_limit}\\
    &0\leq P_{g\text{,}t}\leq\overline{P}_{g}\label{P_grid}\\
    &   \underline{P}_d\leq P_{d\text{,}t}\leq\overline{P}_d\label{DG_output}\\
    &     0\leq P_{{s\text{,}t}}^{+}\text{, } 0\leq P_{{s\text{,}t}}^{-}\label{GES_leq}\\
    &\mathbb{P}(P_{s\text{,}t}^{+}\leq\overline{P}_{s\text{,}t}^{+})\geq1-\epsilon\text{, } \mathbb{P}(P_{s\text{,}t}^{-}\leq\overline{P}_{s\text{,}t}^{-})\geq 1-\epsilon\label{GES_req}\\
    &\mathbb{P}\left(\underline{E}_{s\text{,}t}\leq E_{s\text{,}t}\right)\geq1-\epsilon\text{, }\mathbb{P}\left( E_{s\text{,}t}\leq\overline{E}_{s\text{,}t}\right)\geq1-\epsilon\label{GES_SOC_limit}\\
    & E_{s\text{,}t} =(1-\varepsilon_{s})E_{s\text{,}t-1}+\eta_{s}P_{s\text{,}t}^{-}-P_{s\text{,}t}^{+}/\eta_{s}+\varsigma_{s\text{,}t}\label{GES_SOC}\\
    &E_{s\text{,}T}= E_{s\text{,}0}\label{GES_SOC_cycle}
\end{align}    
\end{subequations}

\noindent where $\mathcal{B}$, $\mathcal{D}$, $\mathcal{S}$, and $\mathcal{T}$ denote the sets of buses, diesel generators, GESs and time periods, respectively, and the subscripts $m/n$, $d$, $s$, and $t$ correspond to the elements within these sets. Decision variables $P_{s\text{,}t}^{+}$/$P_{s\text{,}t}^{-}$, $Q_{s\text{,}t}^{+}$/$Q_{s\text{,}t}^{-}$, and $E_{s\text{,}t}$  denote active discharge/charge power, reactive discharge/charge power and SoC of GES. Decision variables $P_{d\text{,}t}$/$Q_{d\text{,}t}$ and $P_{g\text{,}t}$ denote the active/reactive power outputs from the diesel generator and active power from the main grid, respectively. Decision variables $p_{m\text{,}t}$/$q_{m\text{,}t}$, $P_{nm\text{,}t}$/$Q_{nm\text{,}t}$, $I_{nm\text{,}t}$, $V_{m\text{,}t}$ denotes active/reactive power injection of bus $m$, active/reactive power flow of line $nm$, squared current of line $nm$ and squared voltage of bus $m$, respectively. All the decision variables for power are normalized per time step, and the reactive power is coupled with active power using a fixed power factor. Parameters $c_{s}^{+}$/$c_{s}^{-}$, $c_d$, $c_{g\text{,}t}$ denote operational cost of discharge/charge power of GES, operational cost of diesel generator and electricity price, respectively. Parameters $R_{nm}$, $X_{nm}$, and $Z_{nm}$ denote the resistance, reactance, and impedance of line $nm$, respectively. Parameters $\underline{V}_{m}$, $\overline{V}_{m}$, and $\overline{I}_{nm}$ denote the lower and upper bounds of squared voltage and upper bound of squared current, respectively. Parameters $P_{l\text{,}t}$ and $Q_{l\text{,}t}$ denote the active and reactive power of load, respectively. Parameters $\overline{P}_{g}$, $\overline{P}_d$/$\underline{P}_d$, $\overline{P}_{s\text{,}t}^{+}$/$\overline{P}_{s\text{,}t}^{+}$, $\overline{E}_{s\text{,}t}$/$\underline{E}_{s\text{,}t}$ denote the upper bound of power imported from the main grid, the upper and lower bounds of power output from the diesel generator, the upper bounds of discharge/charge power of GES, and the upper and lower bounds of the SoC, respectively. Parameters $\varepsilon_{s}$, $\eta_{s}$ and $\varsigma_{s\text{,}t}$ denote the self-discharge rate, efficiency, and baseline consumption of GES, respectively. Parameter $\epsilon$ denotes the probabilistic level of chance-constraints.

The objective function~\eqref{objective} minimizes the total operational cost of the microgrid, including the incentive cost, degradation cost, fuel cost of each generation unit, and electricity purchase cost from the main grid. Constraints~\eqref{P_flow}–\eqref{I_limit} define the DistFlow model with second-order cone programming relaxation. Constraints~\eqref{P_flow}-\eqref{Q_flow} enforce the active and reactive power balance. Constraints~\eqref{P_input}-\eqref{Q_input} define the active and reactive power injection. Constraints~\eqref{V_def}-\eqref{SOCP} define the relationship among squared voltage, squared current, active and reactive power flow. Constraints~\eqref{V_limit}-\eqref{I_limit} limit the squared voltage and squared current. Constraint~\eqref{P_grid} limits the power imported from the main grid. Constraints~\eqref{DG_output} limit the power output of the diesel generator. Constraints~\eqref{GES_leq} limit the lower power bounds of GES. Chance-constraints~\eqref{GES_req}-\eqref{GES_SOC_limit} limit the upper power bounds and SoC bounds of GES. The time-varying and stochastic power and SoC bounds of GES can be obtained by data-driven methods (i.e., load decomposition and parameter identification)~\cite{qi2020smart}. Constraint~\eqref{GES_SOC} defines the SoC dynamics with charge/discharge power and additional energy from baseline consumption. Constraints~\eqref{GES_SOC_cycle} enforce the SoC recovery of GES. The complementarity constraint preventing simultaneous charging and discharging of the GES can be relaxed, and we provide a rigorous proof in Appendix~\ref{appendix A}.

The proposed formulation and method are generalized for any time resolution (e.g., 5-minute, 15-minute, 1-hour) and any convex power flow model (e.g., Disflow with SOCP relaxation~\cite{qi2022reliability}, Disflow with semidefinite programming relaxation~\cite{abdi2017review}, linearized Disflow~\cite{li2019temporally}). Considering practical operational requirements for accuracy and computational efficiency, we adopt a 5-minute time resolution and the DistFlow model with SOCP relaxation. The OED problem is practically intractable due to the presence of chance constraints~\eqref{GES_req}-\eqref{GES_SOC_limit} and nonanticipativity constraints~\eqref{P_flow}-\eqref{SOCP}. However, it serves as a theoretical benchmark and motivates the subsequent preliminaries and the proposed solution methodology.

\subsection{Deterministic Reformulation}

To ensure computational tractability, we adopt a deterministic reformulation method to transform the chance constraints into deterministic inequalities~\cite{vrakopoulou2017chance}. Given an assumed probability distribution, with estimated statistical parameters (mean $\mu_{t}$ and standard deviation $\sigma_{t}$), chance-constraints~\eqref{GES_req}-\eqref{GES_SOC_limit} admit a deterministic reformulation in \eqref{reformation}.
\begin{subequations}\label{reformation}
\begin{align}
&P_{s\text{,}t}^{+}\leq \mu_{\overline{P}_{s\text{,}t}^{+}}-F_{\overline{P}_{s\text{,}t}^{+}}^{-1}(1-\epsilon)\sigma_{\overline{P}_{s\text{,}t}^{+}}\\
&P_{s\text{,}t}^{-}\leq\mu_{\overline{P}_{s\text{,}t}^{-}}-F_{\overline{P}_{s\text{,}t}^{-}}^{-1}(1-\epsilon)\sigma_{\overline{P}_{s\text{,}t}^{-}}\\
&E_{s\text{,}t}\leq\mu_{\overline{E}_{s\text{,}t}}-F_{\overline{E}_{s\text{,}t}}^{-1}(1-\epsilon)\sigma_{\overline{E}_{s\text{,}t}}\\
&\mu_{\underline{E}_{s\text{,}t}}+F_{\underline{E}_{s\text{,}t}}^{-1}(1-\epsilon)\sigma_{\underline{E}_{s\text{,}t}}\leq E_{s\text{,}t}
\end{align}
\end{subequations}

\noindent where the normalized inverse cumulative distribution function $F^{-1}$ can be obtained by a known distribution (e.g., Gaussion distribution), distributionally robust approximation, or learned via Maximum Likelihood Estimation~\cite{qi2025locational}.

\subsection{Single-Period Economic Dispatch}

The real-time economic dispatch of a microgrid typically considers either a single-period horizon or a short look-ahead window. We consider a single-period economic dispatch (SED) defined as~\eqref{SED}. Sequentially solving the SED problem is myopic and not equivalent to solving the OED problem, as it overlooks the future opportunity of GES and the time-coupling constraints. Next, we propose a prediction-free solution methodology to effectively address these two issues.
\begin{subequations}\label{SED}
\begin{align}
   &\min \sum_{s\in\mathcal{S}}(c_{s}^{+}P_{s\text{,}t}^{+}+c_{s}^{-}P_{s\text{,}t}^{-})+\sum_{d\in\mathcal{D}}c_dP_{d\text{,}t}+c_{g\text{,}t}P_{g\text{,}t}\label{objective1}\\
    & \text{s.t. }\eqref{P_flow}-\eqref{GES_leq}\text{, }\eqref{reformation}\text{, }\eqref{GES_SOC}
\end{align}    
\end{subequations}

\section{Prediction-Free Two-Stage Coordinated Optimization Framework}\label{method}

\begin{figure}[b!] 
    \setlength{\abovecaptionskip}{-0.1cm}  
    \setlength{\belowcaptionskip}{-0.1cm}   
\centerline{\includegraphics[width=0.95\columnwidth]{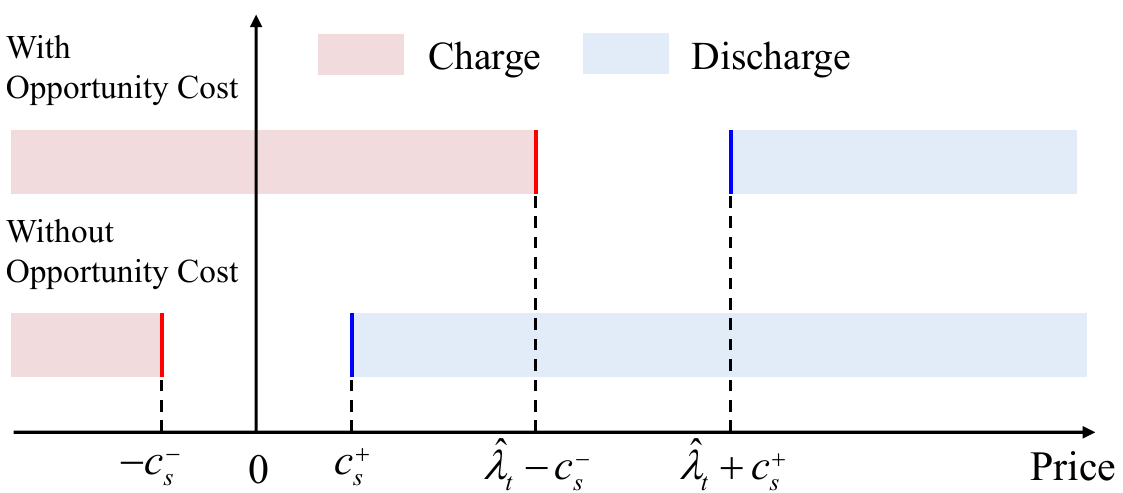}}
    \caption{Illustration of GES actions with and without opportunity cost.}
    \label{influence}
\end{figure}

\subsection{Motivation and Dispatch Framework}

To effectively address the future opportunity value of GES and the time-coupling constraints, we propose the following two promising directions:

\begin{enumerate}
    \item \textbf{Learning the SoC trajectory of GES.} Given a known hindsight SoC trajectory, the SED problem becomes equivalent to the OED problem, since the remaining variables and constraints are all single-period. Therefore, the OED problem can be naturally decomposed into a sequence of SED problems.
    \item \textbf{Learning the opportunity cost of GES.} The decisions of GES depend on the truthful marginal cost of GES and electricity price~\cite{xu2020operational}. After using partial Lagrangian relaxation~\cite{qi2025locational}, the truthful marginal cost is illustrated in~\eqref{eq:truecost}, including both physical degradation cost and future opportunity cost $\hat{\lambda}_t$. The opportunity cost represents the opportunity value of the remaining storage capacity. When the GES operational cost transitions from physical degradation costs to truthful marginal costs, the corresponding GES decision policy undergoes a significant change. As illustrated in Figure~\ref{influence}, without considering future opportunities, storage will discharge whenever the electricity price exceeds the physical discharge cost, and charge when the price drops below the negative physical charging cost. Consequently, the storage almost always favors discharging, given that negative electricity prices are rare. In contrast, when future opportunity costs are considered, the trigger prices become higher, enhancing GES’s ability to capture the arbitrage opportunities and resulting in more economically rational GES actions.
    \begin{align}\label{eq:truecost} \hat{c}_{s}^{+}=c_{s}^{+}+\hat{\lambda}_t\text{, } \hat{c}_{s}^{-}=c_{s}^{-}-\hat{\lambda}_t
    \end{align}
\end{enumerate}

However, both directions encounter challenges. On the one hand, learning the SoC trajectory has demonstrated effectiveness primarily for long-duration energy storage~\cite{qi2025long}, where the trajectory exhibits pronounced seasonal regularity and low sensitivity to electricity prices. In contrast, for short-duration GES with high price sensitivity, prediction-free techniques may yield insufficient accuracy, whereas learning-based or forecast-based methods lack theoretical performance guarantees. On the other hand, directly estimating the opportunity cost or value function is challenging due to its dependence on the SoC and inherent coupling among multiple GES units. Consequently, DP method and its variants have limited applicability in microgrid dispatch with multiple GESs.

\begin{figure*}[t!] 
    \setlength{\abovecaptionskip}{-0.1cm}  
    \setlength{\belowcaptionskip}{-0.1cm}   
    \centerline{\includegraphics[width=1.7\columnwidth]{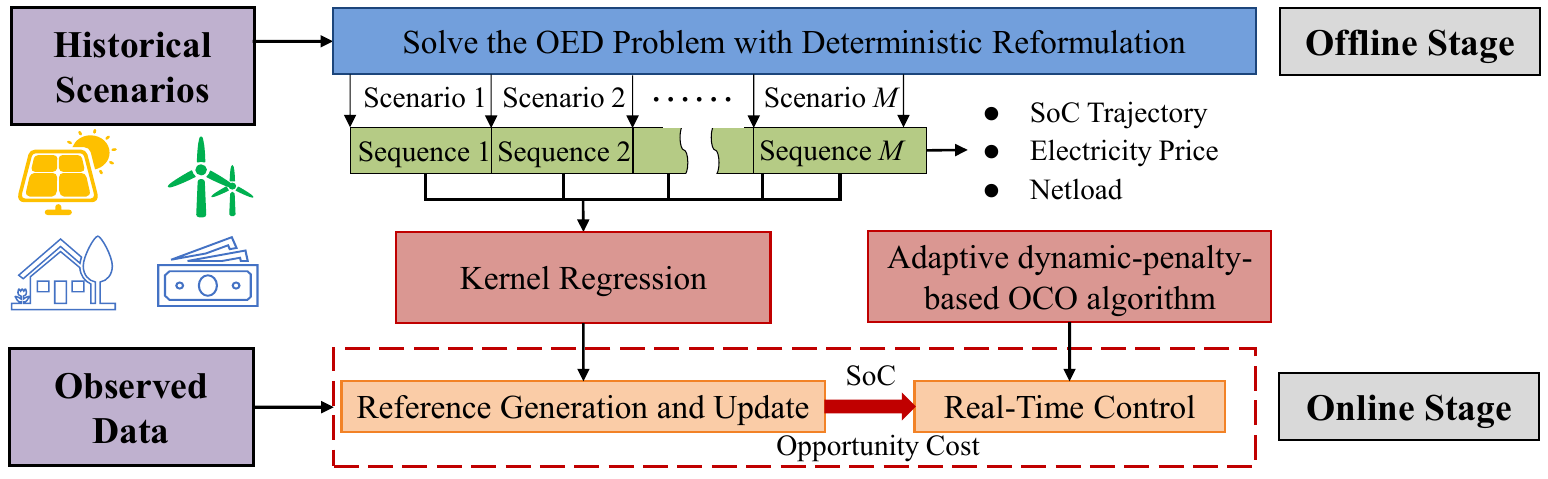}}
    \caption{Prediction-free two-stage coordinated dispatch framework.}
    \captionsetup{justification=centering}
    \label{dispatch framework}
\end{figure*}

Motivated by these challenges, we proposed a prediction-free two-stage coordinated dispatch framework as illustrated in Figure~\ref{dispatch framework}. In the offline stage, we generate sequences of SoC trajectories, prices, and netload for each scenario (day). In the online stage, we first generate and update the reference for SoC trajectory and opportunity cost using kernel regression and offline sequences. Meanwhile, we develop an \rev{adaptive Lagrange multiplier-based} OCO algorithm to track these references without relying on future predictions. The proposed framework utilizes only historical information, thus providing strong robustness in practical microgrid dispatch applications. The detailed procedures are as follows.

\subsection{Offline Stage: Sequences Generation}

We solve the OED problem using the reformulation in~\eqref{reformation} with historical netload and electricity price data to generate the hindsight SoC trajectories $E_{t\text{,}u}$. We collect the sequences for netload, electricity price and SoC trajectory for each scenario, as illustrated in~\eqref{offline1}. These sequences serve as the input data for online reference learning.

\begin{equation}\label{offline1}
\{P_{l\text{,}t\text{,}u}\text{,}\ c_{g\text{,}t\text{,}u}\text{,}\ E_{t\text{,}u}\}_{t=1}^{T}\text{,}\ u\in\{1\text{,}2\text{,}\cdotp\cdotp\cdotp\text{,}M\}
\end{equation}

\noindent where $M$ is the number of historical scenarios.

\subsection{Online Stage: Reference Generation and Update}

We employ the Nadaraya–Watson kernel regression technique~\cite{tsybakov2009nonparametric} to learn the reference for SoC trajectory and opportunity cost. We first define the vectors $P_{l\text{,}[t]}$ and $c_{g\text{,}[t]}$ in~\eqref{online1} to represent the netload and electricity price observed in the real-time operation from the beginning of the operating day to the current period $\textit{t}$. Similarly, we define vectors corresponding to historical scenario $u$ in~\eqref{online2}. Subsequently, we calculate the similarity between real-time observed vectors with historical ones using Euclidean distance and compute the Gaussion kernels to obtain the similarity weight in~\eqref{online3}. $x$ and $y$ denote the input vectors, and $\tau$ represents the bandwidth. Finally, we generate the reference for SoC trajectory and opportunity cost in~\eqref{online4} and \eqref{online5}, respectively. 

The SoC trajectory reference leverages similarities in both price and netload, while the opportunity cost reference utilizes price similarity along with the historical average price $\overline{c}_{g\text{,}u}$. The rationale behind reference learning is straightforward: if the real-time uncertainty realization closely resembles a particular historical day, a higher weight is assigned to that day. The final reference is then obtained as a weighted linear combination of these historical sequences. Moreover, the reference is continuously updated based on the latest observed information, thus indicating more accurate knowledge compared to fixed references, such as historical averages or day-ahead strategies. Compared to learning-based methods, kernel regression offers theoretical performance guarantees and avoids the need for extensive hyper-parameter tuning.
\begin{subequations}\label{kernel_regression}
\begin{align}
    & P_{l\text{,}[t]}=[P_{l\text{,}1}\text{,}\ \cdotp\cdotp\cdotp\text{,}\ P_{l\text{,}t}]\text{,}\ c_{g\text{,}[t]}=[c_{g\text{,}1}\text{,}\ \cdotp\cdotp\cdotp\text{,}\ c_{g\text{,}t}]\label{online1}\\
    & P_{l\text{,}[t]\text{,}u}=[P_{l\text{,}1\text{,}u}\text{,}\ \cdotp\cdotp\cdotp\text{,}\ P_{l\text{,}t\text{,}u}]\text{,}\ c_{g\text{,}[t]\text{,}u}=[c_{g\text{,}1\text{,}u}\text{,}\ \cdotp\cdotp\cdotp\text{,}\ c_{g\text{,}t\text{,}u}] \label{online2}\\
    & K_t(\xi\text{,}\xi')=\exp(-\dfrac{(\|\xi-\xi'\|_2)^2}{t\tau^{2}})\label{online3}\\
    & \hat{E}_{s\text{,}t}=\sum_{u=1}^M\dfrac{K_t(P_{l\text{,}[t]}\text{,} P_{l\text{,}[t]\text{,}u})\cdot K_t(c_{g\text{,}[t]}\text{,} c_{g\text{,}[t]\text{,}u})}{\sum_{u'=1}^{M}K_t(P_{l\text{,}[t]}\text{,} P_{l\text{,}[t]\text{,}u'})\cdot K_t(c_{g\text{,}[t]}\text{,} c_{g\text{,}[t]\text{,}u'})}{E}_{s\text{,}t\text{,}u}\label{online4}\\
    & \hat{\lambda}_t=\sum_{u=1}^M\dfrac{K_t(c_{g\text{,}[t]}\text{,} c_{g\text{,}[t]\text{,}u})}{\sum_{u'=1}^{M}K_t(c_{g\text{,}[t]}\text{,} c_{g\text{,}[t]\text{,}u'})}\overline{c}_{g\text{,}u}\label{online5}
\end{align}
\end{subequations}

\subsection{Online Stage: Real-Time Control Policy}

The real-time dispatch aims to minimize instantaneous operational costs while closely tracking the updated reference, as formulated in~\eqref{tracking}. $\varphi$ is the penalty coefficient of reference tracking. Note that although constraint~\eqref{GES_SOC_cycle} is omitted from the real-time dispatch, continuously tracking the updated reference still ensures a sustainable SoC for GES over time.
\begin{subequations}\label{tracking}
\begin{align}
   &\hspace{-0.1cm}\min \sum_{s\in\mathcal{S}}(\hat{c}_{s}^{+}P_{s\text{,}t}^{+}+\hat{c}_{s}^{-}P_{s\text{,}t}^{-}+\varphi(E_{s\text{,}t}-\hat{E}_{s\text{,}t})^2)+\sum_{d\in\mathcal{D}}c_dP_{d\text{,}t}+c_{g\text{,}t}P_{g\text{,}t}\label{objective2}\\
    & \text{s.t. }\eqref{P_flow}-\eqref{GES_leq}\text{, }\eqref{reformation}\text{, }\eqref{GES_SOC}
\end{align}    
\end{subequations}

Probelm~\eqref{tracking} admits a compact form in~\eqref{compact}. $f_t$, $h_t$, and $x_t$ denote the time-varying objective function, constraints and decision variables, respectively. Equality constraints can be converted into pairs of inequality constraints and included in $h_t(x_t)\leq 0$. We should note that~\eqref{compact} cannot be solved directly without observing uncertainties at the current time step. To eliminate reliance on future predictions and implement a ``0-look-ahead" decision pattern, we propose an \rev{adaptive Lagrange multiplier-based} OCO algorithm. We notice that the performance of OCO is inherently sensitive to the stepsize, hence we adopt the adaptive expert-tracking framework~\cite{zhang2018adaptive}, where each expert $i$ operates with distinct stepsizes $\alpha_{i\text{,}t-1}$ \text{and} $\beta_{i\text{,}t-1}$. The \rev{adaptive Lagrange multipliers} $\nu_{i\text{,}t}$ and decisions $x_{i\text{,}t}$ are updated in~\eqref{Q_update} and~\eqref{x_update} for each expert in parallel, respectively. $\left\langle \Gamma\text{,}\Gamma'\right\rangle$ denotes the standard inner product of two vector $\Gamma$ and $\Gamma'$. $X$ denotes the feasible sets. Subsequently, we compute the surrogate loss $\ell_{i\text{,}t}$ using~\eqref{loss}. Finally, the weights $\rho_{i,t}$ for each expert are updated according to its empirical performance measured by surrogate losses using~\eqref{final_decision}, and the final decision is derived as the weighted average of all experts' decisions.
\begin{subequations}\label{OCO}
\begin{align}
& \min \ f_t(x_t)\quad \text{s.t. } h_t(x_t)\leq0,\mathrm{~}t=1,2,...,T  \label{compact}\\
&\nu_{i\text{,}t-1}=\max(\nu_{i\text{,}t-2}+\beta_{t-1}[h_{t-1}(x_{t-1})]_+\text{,}\ \theta_{i\text{,}t-1})\label{Q_update}\\
& x_{i\text{,}t}=\arg\min_{x\in X}\{\alpha_{i\text{,}t-1}\left\langle\partial f_{t-1}(x_{i\text{,}t-1})\text{,}\ x-x_{i\text{,}t-1}\right\rangle\label{x_update}\\
&\hspace{0.5cm}+\alpha_{i\text{,}t-1}\beta_{t-1}\left\langle \nu_{i\text{,}t-1}\text{,}\ [h_{t-1}(x)]_+\right\rangle+\|x-x_{i\text{,}t-1}\|^2\}\notag \\
& \ell_{i\text{,}t-1}=\langle\partial {f}_{t-1}({x}_{t-1})\text{,}\ {x}_{i\text{,}t-1}-{x}_{t-1}\rangle\label{loss}\\
& \rho_{i\text{,}t}=\frac{\rho_{i\text{,}t-1}e^{-\gamma \ell_{i\text{,}t-1}}}{\sum_{i=1}^N\rho_{i\text{,}t-1}e^{-\gamma \ell_{i\text{,}t-1}}}\text{, }{x}_t=\sum\nolimits_{i=1}^N\rho_{i\text{,}t}{x}_{i\text{,}t}\label{final_decision}
\end{align}
\end{subequations}

\begin{remark}\textbf{\emph{Rationale.}}
 The key idea of the proposed algorithm is to leverage information from the previous time step to approximate the current system state, while ensuring feasibility via \rev{adaptive Lagrange multipliers} for constraint violations. Specially, ${f}_{t}({x})$ and ${h}_t({x})$ are approximated using the first-order Taylor expansion $\left\langle\partial f_{t-1}(x_{i\text{,}t-1})\text{,}\ x-x_{i\text{,}t-1}\right\rangle$ and clipped constraint function $\left[{h}_{t-1}({x})\right]_{+}$. \rev{Adaptive Lagrange multipliers} $\nu_{i\text{,}t}$ substitute dual variables for constraints $h_t$ and is increased when constraints are violated, thereby dynamically adjusting the penalty for constraint violations. Compared to the existing OCO framework, we generate \rev{adaptive Lagrange multipliers} only for nonanticipativity constraints~\eqref{P_flow}–\eqref{P_grid} and apply Lagrangian relaxation to penalize constraint violations caused by the information gap. Other deterministic box constraints \eqref{DG_output}, \eqref{GES_leq} and \eqref{reformation} form the feasible set X, rendering the projection~\eqref{x_update} computationally trivial and easily solved in real-time. We also impose a lower bound on the \rev{adaptive Lagrange multipliers} by introducing $\theta_{i\text{,}t-1}$ to prevent the algorithm from taking aggressive decisions that could lead to large constraint violations. Bregman divergence $\|x-x_{i\text{,}t-1}\|^2$ is introduced to smooth the difference between coherent actions and enhance the stability of the algorithm. Moreover, instead of using a fixed learning rate, we incorporate an expert-tracking mechanism and the weighting mechanism adaptively favors experts with superior historical performance, thus enhancing the responsiveness and adaptability of the algorithm.  
\end{remark}

\begin{theorem}\label{tm}\textbf{\emph{Sublinear Bounds for Dynamic Regret and Time-Varying Hard Constraint Violations.}} Suppose all the assumptions in~\eqref{assump1}-\eqref{assump3} hold, given parameters setting in~\eqref{parameter}, we can achieve the sublinear bounds of dynamic regret and time-varying hard constraints violation in~\eqref{main_result}.

Assumption 1: $f_{t}$ and $h_{t}$ are convex functions. The set \textit{X} is convex, and there is a positive constant W such that:
\begin{equation}\label{assump1}
\parallel x-y\parallel\leq W\text{,}\ \forall x\text{,}y\in X
\end{equation}

Assumption 2: There exists a positive constant 
\textit{F} such that:
\begin{equation}\label{assump2}
\mid f_t(x)-f_t(y)\mid\leq F\text{,}\ \parallel h_t(x)\parallel\leq F\text{,}\ \forall t\text{,}\ \forall x\text{,}y\in X
\end{equation}

Assumption 3: The subgradients $\partial f_t$ and $\partial h_t$ exist. There exists a positive constant \textit{J} such that:
\begin{equation}\label{assump3}
\parallel\partial f_t(x)\parallel\leq J\text{,}\ \parallel\partial h_t(x)\parallel\leq J\text{,}\ \forall t\text{,}\ \forall x\text{,}y\in X
\end{equation}
\begin{equation}\label{parameter}
\begin{aligned}
&N=[\frac{1}{2}\log_{2}(1+T)]+1\text{,}\ \gamma=\frac{1}{\sqrt{T}}\text{,}\ \alpha_{i,t}=\frac{2^{i-1}}{t^{\frac{1}{2}+\chi}}\text{,}\\&\beta_{t}=t^{\frac{1}{2}+\delta}\text{,}\ \theta_{i\text{,}t}=2^{i-1}t\text{,}\ \forall i\in\{1\text{,}2\text{,}\cdotp\cdotp\cdotp\text{,}N\}\text{,}\ \frac{1}{2}>\delta>\chi>0
\end{aligned}
\end{equation}
\begin{subequations}\label{main_result}
    \begin{align}
&\text{Regret}=\sum_{t=1}^T{[f_t(x_t)-f_t(x_t^*)]}=O(T^{\frac12+\chi}\sqrt{1+P_x})\label{regret} \\
&\text{Vio}=\sum_{t=1}^T\lVert[h_t(x_t)]_+\rVert=O(\log_2(T)T^{1-\frac\chi2})\label{vio}
    \end{align}
\end{subequations}
\textcolor{blue}{where $\{x_t^*\}_{t=1}^T$ is the comparator sequence obtained by solving the problem}

\textcolor{blue}{$(x_1^*,\ldots,x_T^*)=\arg\min\limits_{x_1,\ldots,x_T\in X}\ \sum_{t=1}^T f_t(x_t)\ \ \text{s.t. } h_t(x_t)\le 0,\ \forall t$
and $P_x=\sum_{t=1}^{T-1}\parallel x_{t+1}^*-x_t^*\parallel$.}

\end{theorem}

\begin{proof}
\textcolor{blue}{We defer the complete and rigorous proof to Appendix~\ref{appendix B}-\ref{appendix C}.}
\end{proof}

\begin{remark}
\textcolor{blue}{We note that the OCO algorithm is specifically designed to achieve sublinear bounds on both regret and constraint violations. The sublinear performance typically refers to time horizon $T$. To the best of our knowledge, all existing OCO algorithms assume that $P_x$ is independent of $T$. Although accurately estimating $P_x$ remains challenging, achieving overall sublinear performance requires $P_x = o(T^{1-2\chi})$, which can be empirically verified using hindsight results. Furthermore, the proposed algorithm outperforms~\cite{guo2022online,ding2021dynamic}, whose regret scales linearly with $P_x$. Compared to our previous work~\cite{qi2025long}, which only focuses on isolated microgrid operation and dynamic regret, this paper innovatively incorporates the tracking of opportunity prices and redesigns the OCO algorithm to address constraint violations. We note that the proposed OCO algorithm first achieves sublinear bounds for both dynamic regret and time-varying hard constraint violations. In contrast, existing OCO algorithms achieve sublinear bounds either only for regret or partially (e.g., static regret, soft constraint violations).}
\end{remark}

We summarize the proposed prediction-free two-stage coordinated dispatch framework in \textbf{Algorithm~\ref{alg1}}. 

\begin{algorithm}[!t]
\caption{\textbf{:} Prediction-free two-stage coordinated dispatch}\label{alg1}
\begin{algorithmic}
\STATE 
\STATE \textbf{[Offline Stage]}
\STATE \textbf{Input:} Historical scenarios of netload and electricity price $\{P_{l\text{,}t\text{,}u}\text{,}\ c_{g\text{,}t\text{,}u}\}_{t=1}^{T}\text{,}\ u\in\{1\text{,}2\text{,}\cdotp\cdotp\cdotp\text{,}M\}$
\STATE \textbf{Output:} Hindsight SoC trajectories of historical scenarios $\{ E_{t\text{,}u}\}_{t=1}^{T}\text{,}\ u\in\{1\text{,}2\text{,}\cdotp\cdotp\cdotp\text{,}M\}$.
\STATE \textbf{For} $u=1\text{,}\cdots\text{,}M$
\STATE \hspace{0.5cm}Solve the OED problem~\eqref{OED} with reformulation~\eqref{reformation} to
\STATE \hspace{0.5cm}generate hindsight SoC trajectory $E_{t\text{,}u}$.
\STATE \textbf{end}
\STATE \textbf{[Online Stage]}
\STATE \textbf{Input:} Sequences $\{P_{l\text{,}t\text{,}u}\text{,}\ c_{g\text{,}t\text{,}u}\text{,}\ E_{t\text{,}u}\}_{t=1}^{T}\text{,}\ u\in\{1\text{,}2\text{,}\cdotp\cdotp\cdotp\text{,}M\}$ and penalty coefficient $\varphi$.
\STATE \textbf{Output:}  Decisions $x_t$, objective $f_t(x_t)$, Regret and Vio.
\STATE \textbf{Step 1 - Initialization:} 
\STATE \hspace{0.5cm}Set $\nu_{i\text{,}0}=0\text{,}\ x_{i,1}\in X\text{,}\ x_1=\sum_{i=1}^N\rho_{i,1}x_{i,1}$\text{,}
\STATE \hspace{0.5cm}$\rho_{i,1}=(N+1)/[i(i+1)N]\text{,}\ \forall i\in\{1,2\text{,}\cdotp\cdotp\cdotp\text{,}N\}.$
\STATE \textbf{Step 2 - Reference Update and Real-Time Control:}  
\STATE  \textbf{For} $t=2\text{,}\cdots\text{,}T$
\STATE \hspace{0.5cm}Update reference for SoC and opportunity cost via~\eqref{kernel_regression}.
\STATE \hspace{0.5cm}Update \rev{adaptive Lagrange multipliers} via~\eqref{Q_update}.
\STATE \hspace{0.5cm}Update decisions via~\eqref{x_update}.
\STATE \hspace{0.5cm}Compute urrogate loss in parallel via~\eqref{loss}.
\STATE \hspace{0.5cm}Update the expert weight and finalize decisions via~\eqref{final_decision}.
\STATE \textbf{end}
\end{algorithmic}
\end{algorithm}

\subsection{Extensions of the Proposed Framework}

In this subsection, we discuss how to extend our proposed framework to more complex and practical application scenarios.

\subsubsection{Non-Convexity} The proposed framework primarily addresses convex models. Nevertheless, it can also be extended to handle non-convexities typically encountered in practical microgrid operations. For instance, binary variables related to the on-off control of thermostatically controlled loads can be approximated through regularization techniques, while maintaining theoretical performance guarantees, as demonstrated in~\cite{lesage2021online}. Furthermore, other types of non-convexity (e.g., hydrogen efficiency model~\cite{qi2025long}, battery degradation model~\cite{xu2017factoring}) can be handled by employing standard convex approximations (e.g., convex hull) or piecewise-linear approximation, both commonly utilized in practice to ensure computational efficiency and model tractability. Additionally, extensive studies are focusing on convexifying non-convex power flow constraints~\cite{madani2014convex,farivar2013branch,abdi2017review}. Thus, the proposed framework remains flexible and robust to accommodate realistic scenarios involving non-convexities.

\subsubsection{Abnormal State} In practical microgrid operations, faults in lines or units are inevitable. To manage such emergency conditions, one approach is to regenerate the offline sequences based on the current system configuration, which typically requires only a few minutes through parallel computing. Subsequently, references for SoC and opportunity cost can be regenerated for online tracking. Alternatively, microgrids generally maintain reserves or adopt emergency dispatch strategies (e.g., topology reconfiguration~\cite{zhu2025adversarial}, re-dispatch~\cite{abdelmalak2022proactive}). In this regard, the proposed dispatch framework can naturally transition into emergency operation modes and subsequently revert to normal dispatch mode once the fault is cleared. Thus, our proposed framework is applicable to both normal and abnormal states.

\section{Case Study}\label{case study} 

\subsection{Setups}
\begin{figure}[t!] 
    \setlength{\abovecaptionskip}{-0.1cm}  
    \setlength{\belowcaptionskip}{-0.1cm}   
    \centerline{\includegraphics[width=1\columnwidth]{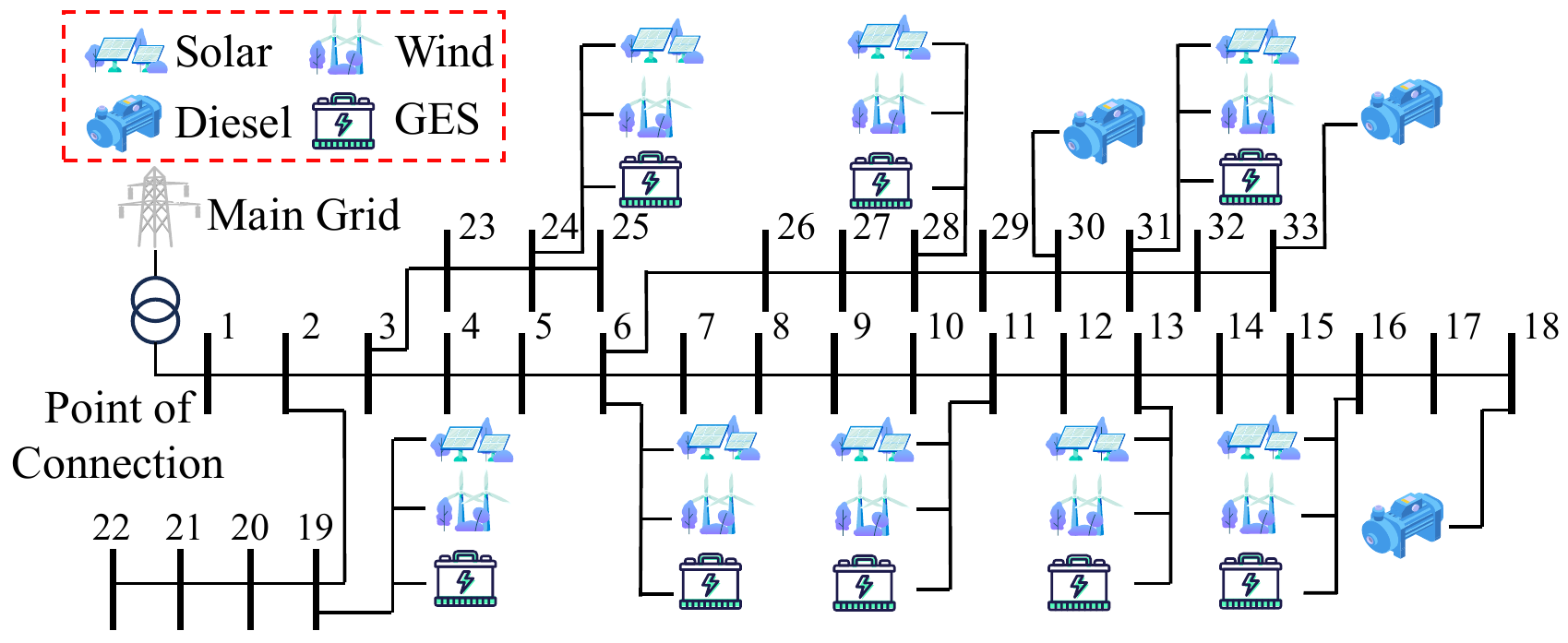}}
    \caption{Diagram of the modified 33-bus radial microgrid system.}
    \label{IEEE33 diagram}
\end{figure}

We test the effectiveness of the proposed framework on the IEEE 33-bus radial system, configured as a microgrid~\cite{wang2014robust}. The test system diagram is shown in Figure~\ref{IEEE33 diagram}, and includes wind generation, solar generation, diesel generation, GES, and local load, with total capacities of 2.5 MW, 2.5 MW, 1.5 MW, 1.8 MW, and 2.5 MW, respectively. Within the GES, the capacities of physical energy storage and virtual energy storage are configured in a 2:1 ratio. The ground-truth data for wind generation, solar generation, load, and electricity prices at a 5-minute resolution are obtained from publicly available datasets provided by the Australian Energy Market Operator. All case study data, including system configurations, model parameters, and algorithm settings, are made publicly available~\cite{huangdata}.

The optimization is coded in MATLAB equipped with the YALMIP interface and solved by the Gurobi 11.0 solver. The programming environment is Intel Core i7-1165G7 @ 2.80GHz with RAM 16 GB.

\subsection{Effectiveness Analysis of the Proposed Method}

\begin{figure}[t!] 
    \setlength{\abovecaptionskip}{-0.1cm}  
    \setlength{\belowcaptionskip}{-0.1cm}   
\centerline{\includegraphics[width=1\columnwidth]{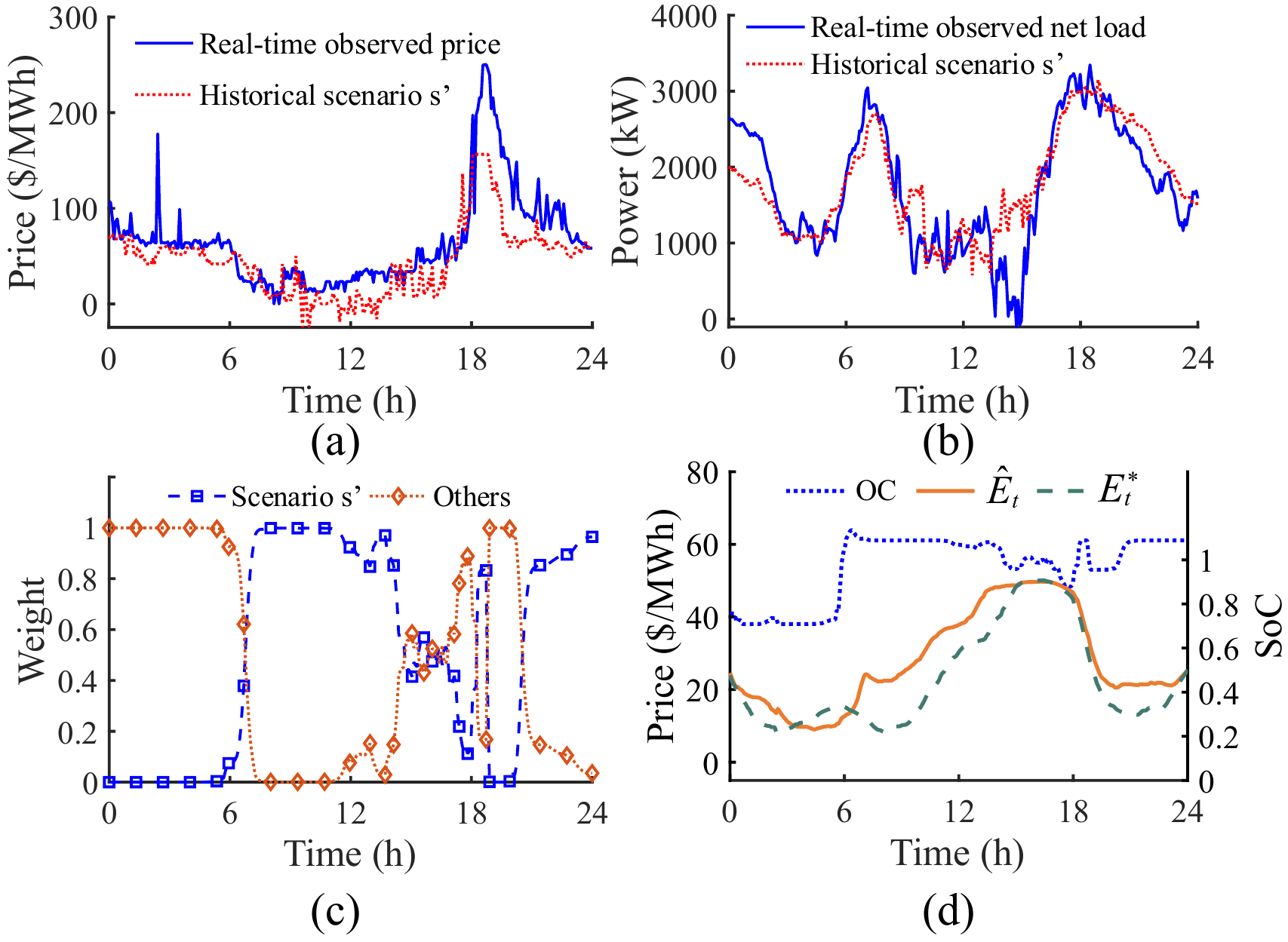}}
    \caption{Results for day 7: (a) prices, (b) netload, (c) weights of historical scenarios, and (d) references (OC: opportunity cost).}
    \label{DAP_update}
\end{figure}

\begin{figure}[t!] 
    \setlength{\abovecaptionskip}{-0.1cm}  
    \setlength{\belowcaptionskip}{-0.1cm}   
\centerline{\includegraphics[width=0.95\columnwidth]{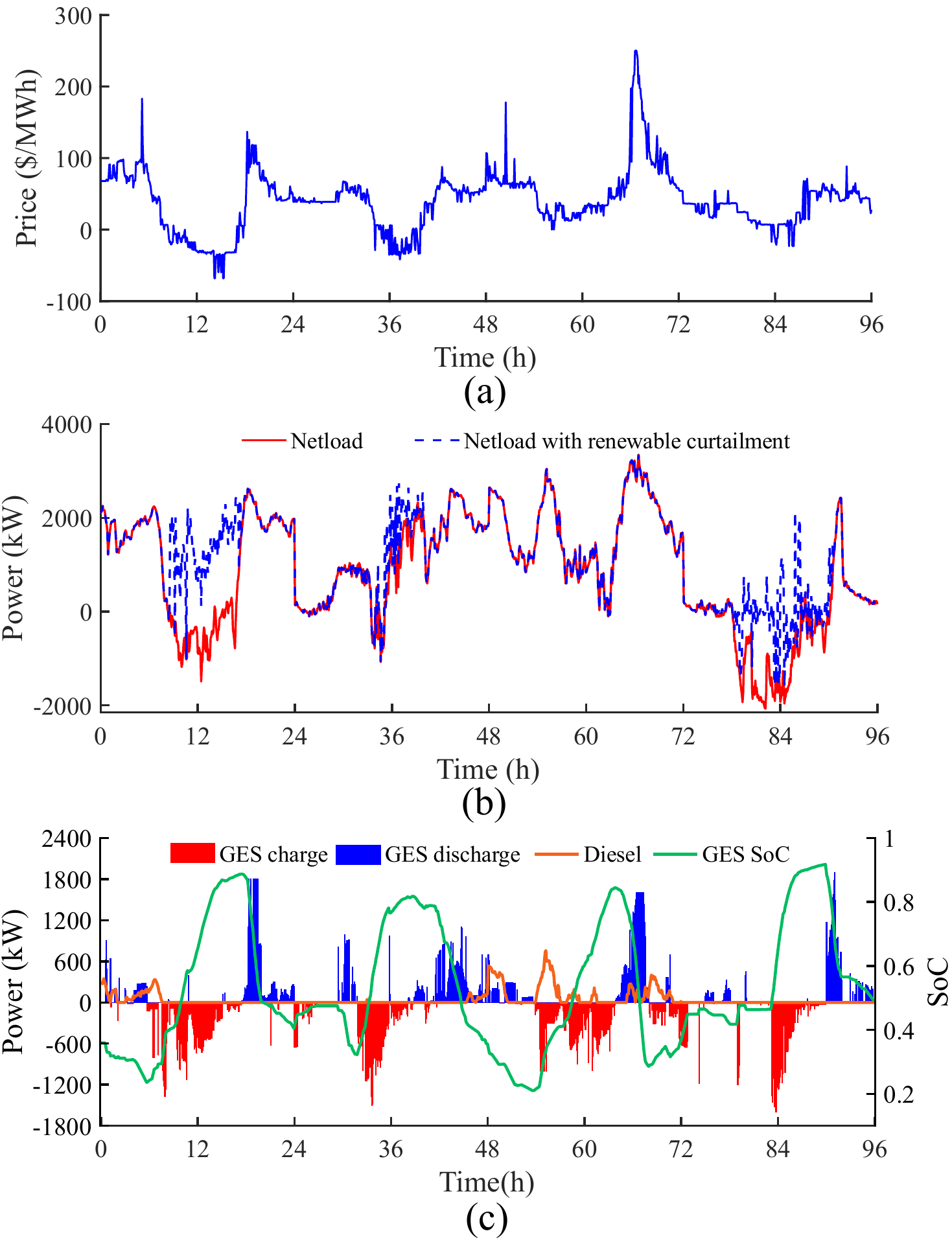}}
    \caption{Results for day 5-8: (a) prices, (b) netload, and (c) dispatch decisions.}
    \label{proposed}
\end{figure}

We implement the proposed dispatch framework over a two-month period to test the average performance. Taking day 7 as an example, Figure~\ref{DAP_update} illustrates the input scenarios, observed scenarios, and dynamic updated reference for SoC and opportunity cost. Scenario $\text{s'}$ denotes the historical scenario most similar to the 7th day, and $E^*_t$ represents the hindsight SoC trajectory of the 7th day (SoC are normalized). The weights assigned to historical scenarios are continuously updated according to their similarity to the current operating day. Moreover, the trend of the SoC reference closely aligns with the hindsight SoC trajectory. The opportunity cost reference fluctuates around \$60/MWh, closely matching the day's actual average opportunity cost of \$65/MWh. Hence, the references generated via kernel regression provide an effective global guidance for online optimization.

\begin{figure}[t!] 
    \setlength{\abovecaptionskip}{-0.1cm}  
    \setlength{\belowcaptionskip}{-0.1cm}   
    \centerline{\includegraphics[width=1\columnwidth]{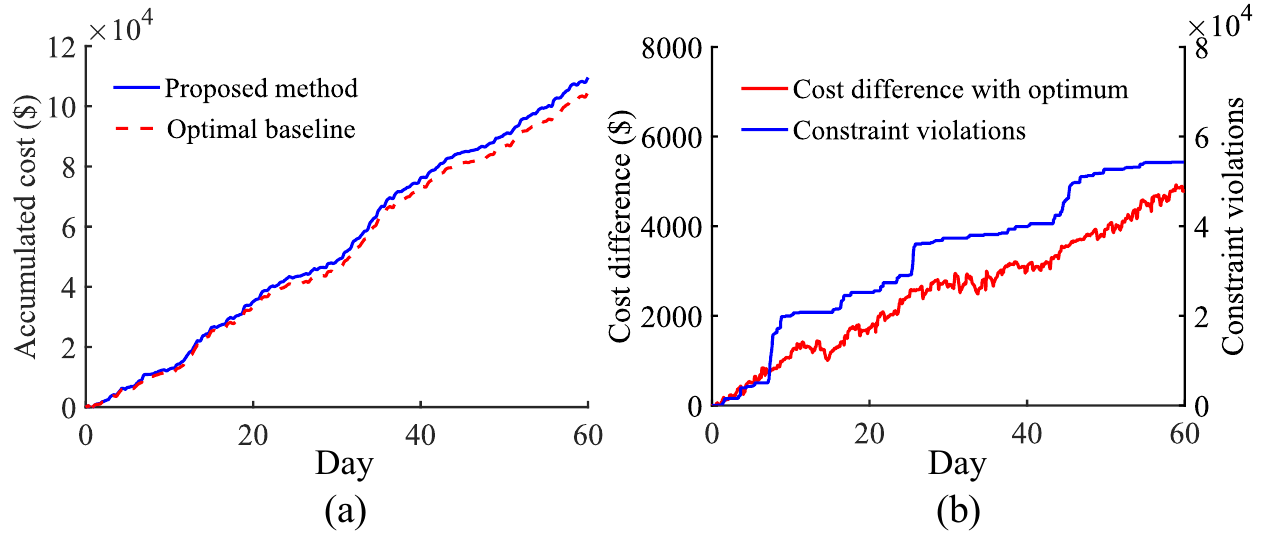}}
    \caption{Performance over 2 months: (a) accumulated cost and (b) cost difference with optimum \& constraint violations.}
    \label{regret-figure}
\end{figure}
Figure~\ref{proposed} illustrates the dispatch results from day 5 to day 8, demonstrating that the real-time dispatch decisions generated by the proposed OCO algorithm effectively minimized operational costs while ensuring grid awareness. Wind and solar generation are curtailed during periods of negative prices. GES dynamically discharges or charges when prices exceed or fall below the truthful marginal costs, respectively, which incorporate future opportunity costs. As shown in Figure~\ref{regret-figure}, the accumulated cost over the two-month period deviated by only 4.57\% from the optimal baseline, demonstrating the effectiveness of the proposed method.  Voltage security is strictly satisfied for 98.24\% of the time. We note that the proposed method cannot entirely eliminate myopia. The temporary decreases in the cost difference curve at certain periods occur because the optimal baseline, which has perfect foresight, may strategically choose higher immediate costs at certain intervals (e.g., by charging GES) to achieve greater long-term economic benefits, especially during subsequent price peaks. In contrast, the proposed method, lacking perfect future information, might occasionally opt for decisions that temporarily yield lower immediate costs at these intervals (e.g., discharging GES). This local phenomenon does not indicate superior overall performance, and the cumulative cost difference remains non-negative, exhibiting a general upward trend over the long term. Additionally, constraint violations are inherently caused by the ``0-lookahead" pattern. In practical operations, the system operator typically either tolerates minor voltage violations to maintain critical loads or resorts to load shedding to strictly ensure voltage security.

\subsection{Comparative Results with Different Dispatch Methods}

We further compare the performance of our method with state-of-the-art methods, as detailed below:

\textbf{(M0) Perfect Foresight:} M0 solves the OED problem in~\eqref{OED} with perfect knowledge of uncertainties. Although impractical in reality, it serves as an optimal baseline for comparison.

 \textbf{(M1) Proposed Method:} M1 is the proposed prediction-free two-stage coordinated dispatch presented in Algorithm~\ref{alg1}.

 \textbf{(M2) MPC-1:} MPC-1, proposed in~\cite{guo2023long}, also employs a two-stage coordinated dispatch framework. However, compared to \textbf{M1}, MPC-1 tracks only the SoC reference and utilizes a rolling-horizon dispatch based on recently updated forecasts. The real-time forecast errors are simulated with a mean absolute percentage error (MAPE) of 10\%. \rev{The look-ahead window is set to 4 hours.}

 \textbf{(M2*) MPC-2:} MPC-2 follows the identical MPC approach as \textbf{M2}, but with higher prediction errors (MAPE of 20\%).

 \textbf{(M3) Lyapunov Optimization-1:} Lyapunov optimization-1 adopts the frameworks proposed in~\cite{stai2020online,shi2015real}, employing a Lyapunov drift-plus-penalty approach to minimize instantaneous operational costs while constraining the SoC within an ideal operational range. The SoC reference (same as \textbf{M1}) is included in the objective function for real-time tracking.

\textbf{(M3*) Lyapunov Optimization-2:} Lyapunov Optimization -2 follows the identical Lyapunov approach as \textbf{M3}, but uses a day-ahead SoC reference generated by scenario-based stochastic optimization~\cite{stai2022computing,stai2017dispatching,su2013stochastic}. The reference is static and will not update in real-time.

\begin{table}[t]
    \setlength{\abovecaptionskip}{-0.1cm}
    \setlength{\belowcaptionskip}{-0.1cm}
    \renewcommand\arraystretch{1.5} 
    \setlength{\tabcolsep}{0.01cm} 
    \caption{Comparisons of Operational Performance over 2 Months}\label{hhh}
    \begin{center}
    \begin{tabular}{c cccccc}
    \toprule
    \multirow{2}{*}{Indices} & \multicolumn{6}{c}{Methods}\\
    \cmidrule(lr){2-7}
     & M0 & M1 & M2 & M2* & M3 & M3*\\ 
    \midrule
    Operation Cost (\$) & 110210 & 115234 & 121346 & 130490 & 122843 & 125518\\
    Voltage Satisfaction Rate (\%) & 100 &98.62 & 89.56 & 87.35 & 97.87 & 97.84\\
    Daily Computation Time (min) & / &2.3 & 46.4 & 46.4 & 2.3& 2.3\\
    \bottomrule  
    \end{tabular}
    \end{center}
\end{table}

\begin{figure}[t!] 
    \setlength{\abovecaptionskip}{-0.1cm}  
    \setlength{\belowcaptionskip}{-0.1cm}   
    \centerline{\includegraphics[width=0.95\columnwidth]{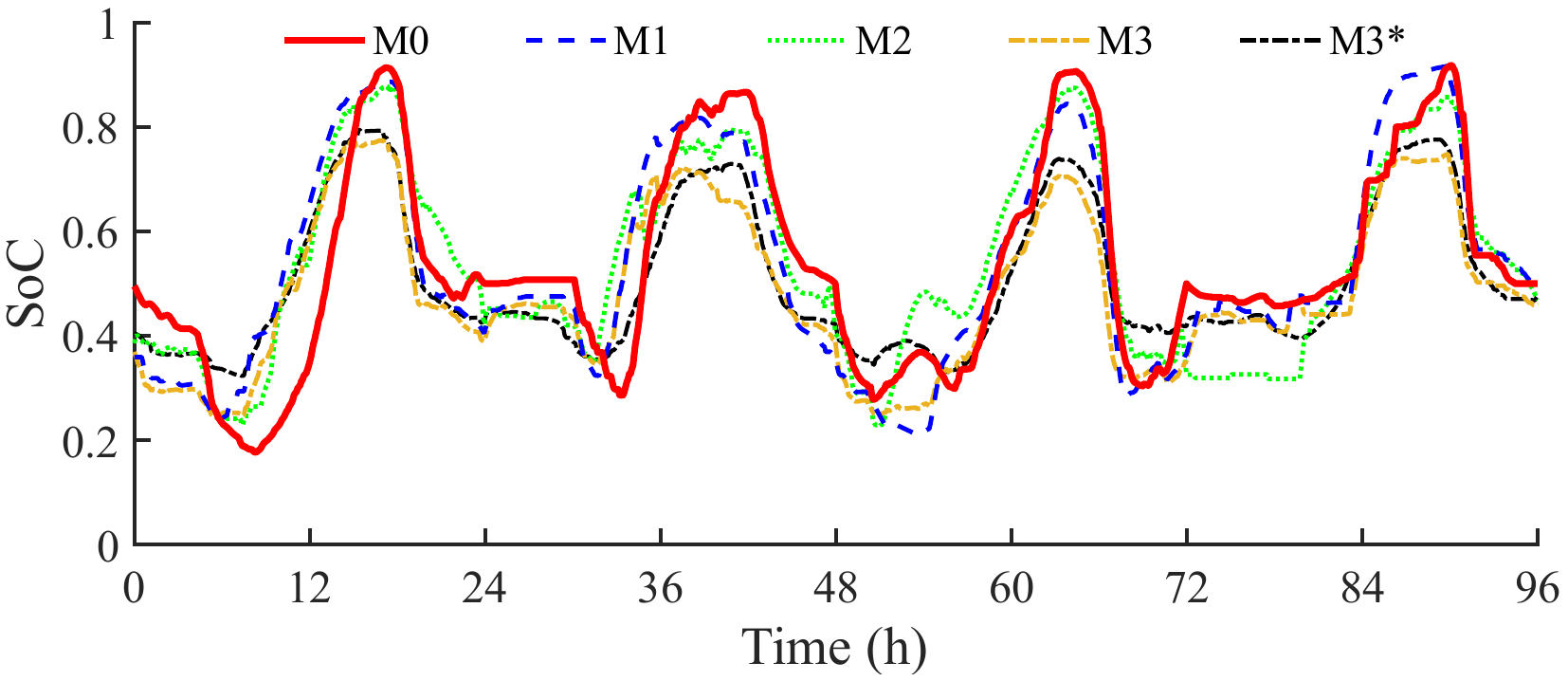}}
    \caption{Comparison of SoC trajectories under different methods for days 5–8.}
    \label{SOC_comparison}  
\end{figure}

\begin{figure}[t!] 
    \setlength{\abovecaptionskip}{-0.1cm}  
    \setlength{\belowcaptionskip}{-0.1cm}   
\centerline{\includegraphics[width=0.95\columnwidth]{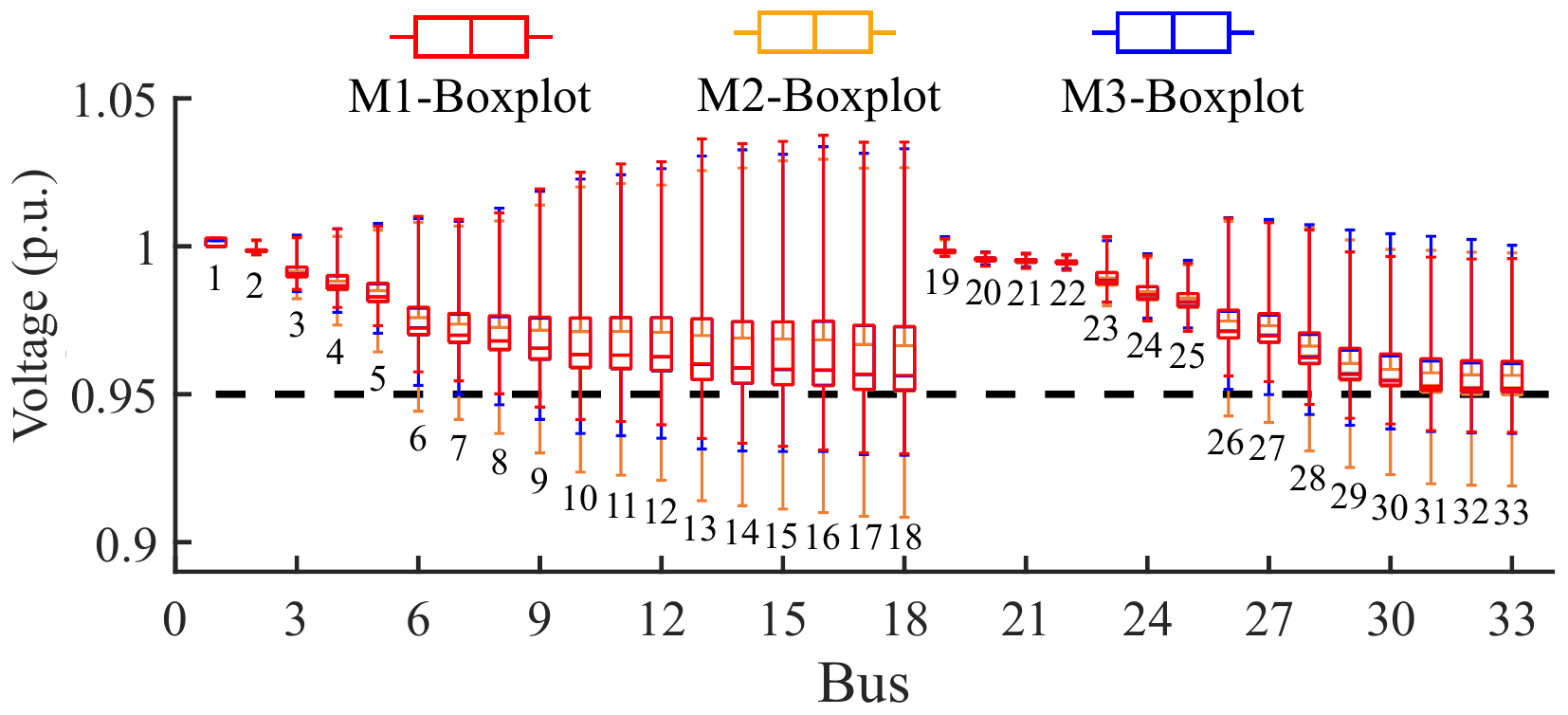}}
    \caption{Comparison of voltage distribution between M1-M3 for day 5-8.}
    \label{voltage distribution}
\end{figure}

The dispatch performance of the six methods over the two-month period is summarized in Table~\ref{hhh}. Figure~\ref{SOC_comparison} and figure~\ref{voltage distribution} compares the SoC trajectories and
 voltage distribution of different methods for days 5 to 8. It is evident that, compared to M1 and M2, the Lyapunov-based dispatch methods (M3 and M3*) deviate more significantly from the optimal baseline (M0). Particularly, the SoC trajectories of M3 and M3* exhibit noticeably conservative behaviors. For example, during periods such as 16–18h and 86–90h, the SoC of M0, M1, and M2 methods are approximately 0.9, whereas those of M3 and M3* remain below 0.8. This conservative behavior inherently arises from the Lyapunov drift-plus-penalty function, which penalizes the deviation of a virtual energy queue from a predefined stable operating range. The Lyapunov drift term explicitly aims at stabilizing the GES state by bounding its cumulative deviation from an ideal range. Consequently, the GES is restricted from performing aggressive charging or discharging to capitalize on large market price differentials, thereby limiting its arbitrage capability. Such inherent conservatism ensures long-term operational stability and reliability, but inevitably compromises economic performance in volatile market environments. This limitation is also reflected in Table~\ref{hhh}, which shows higher operational costs for M3 and M3*. Furthermore, M3* exhibits a higher operational cost compared to M3, primarily because M3 dynamically updates the SoC reference leveraging real-time information, while M3* tracks a static, day-ahead SoC trajectory. Hence, M3* inevitably leads to lower economic performance.

Furthermore, although the SoC trajectory of M2 generally resembles those of M1 and M0, inaccurate forecasts or references may lead to suboptimal decisions, as observed during hours 19–24, 34–41, 50–56, 66–69, and 83–86. MPC-based methods exhibit declined performance with increasing prediction errors. This is the inherent limitation of prediction-dependent methods. In contrast, M1 achieves better overall performance because tracking both the SoC trajectory and opportunity cost provides GES decisions with a global perspective, thus mitigating myopic behaviors. Additionally, the OCO algorithm can quickly respond to changes in netload and prices while continuously adapting its learning rate. Under extreme scenarios, such as the unexpected price surge during hour 19, M1 promptly increased discharge power, significantly reducing operational costs compared to M2. Similarly, during negative price spikes at hour 85, M1 responded swiftly by increasing charging power, again reducing operational costs. Additionally, during hours 64–68, as netload peaked rapidly, M1 effectively prevented voltage violations by timely adjusting discharge power, whereas M2 caused excessive voltage drops due to inadequate response, as shown in Figure~\ref{voltage distribution}. Overall, M1 demonstrates superior adaptability to time-varying conditions.

In terms of computational efficiency, M1 and M3 significantly outperform M2, as both M1 and M3 are prediction-free and involve only single-period optimization problems at each time step. In contrast, M2 requires optimization over look-ahead windows, thus reducing its computational efficiency.

To further demonstrate the advantages of the proposed dynamically updated SoC reference, we tested three online optimization methods (OCO, MPC, and Lyapunov optimization) under three distinct reference scenarios: (1) dynamically updated SoC reference; (2) day-ahead SoC reference; (3) no reference. The comparative results are summarized in Table~\ref{reference_comparison}. The proposed OCO method consistently achieves the lowest operational cost across all three references, demonstrating its superior performance in real-time dispatch. Furthermore, dispatch methods without any reference tracking result in myopic decisions and the highest operational costs, which hindsight the importance of reference tracking. 
Additionally, dynamically updating the SoC reference based on observed uncertainties significantly outperforms the static day-ahead reference, as our approach adaptively adjusts to real-time fluctuations in RES generation and market prices, thereby enhancing the microgrid's operational flexibility. In contrast, the static day-ahead plan cannot adapt to real-time deviations.

\begin{table}[h]
    \setlength{\abovecaptionskip}{-0.1cm}
    \setlength{\belowcaptionskip}{-0.1cm}
    \renewcommand\arraystretch{1.5} 
    \setlength{\tabcolsep}{0.25cm} 
    \caption{{Cost Comparison (\$) under Different References and Methods}}\label{reference_comparison}
    \begin{center}
    \begin{tabular}{c c c c}
    \toprule 
    \multirow{2}{*}{References} & \multicolumn{3}{c}{Methods}\\
    \cmidrule(lr){2-4}
    & {OCO} & {MPC} & {Lyapunov} \\ 
    \midrule
    {Dynamically Updated Reference} & {115234} & {121346} & {122843} \\
    {Day-ahead Reference} & {120418} & {122992} & {125516}  \\
    {No Reference} &  {123298} & {126890} & {129375} \\
    \bottomrule  
    \end{tabular}
    \end{center}
\end{table}

\subsection{Sensitivity Analysis}

\subsubsection{Result sensitivity to reference tracking}

First, we test the proposed method with different penalty coefficients for reference tracking in Figure~\ref{weight_of_SOC}. We observe that the cost performance is sensitive to penalty coefficients, initially increasing and then decreasing as the penalty coefficient increases. An interesting observation is that strictly following the reference yields worse performance than having no reference at all. This indicates that rigid adherence to the reference reduces flexibility and adaptability in volatile environments.

\begin{table}[t!]
    \setlength{\abovecaptionskip}{-0.1cm}
    \setlength{\belowcaptionskip}{-0.1cm}
    \renewcommand\arraystretch{1.5} 
    \setlength{\tabcolsep}{0.4cm} 
\caption{Comparisons Between Different Variants of M1}\label{different steps}
    \begin{center}
    \begin{tabular}{ccccccc}
    \toprule 
    Indices & M1 & M1-a &M1-b&M1-c \\ 
    \midrule
    Cost (\$) & 115234 & 144043 & 123298 & 119961\\
    \bottomrule  
    \end{tabular}
    \end{center}
\end{table}

Furthermore, we investigate the performance of several variants of the proposed method, as detailed below:

\textbf{(1) M1-a:} Large penalty ($\varphi=1000$) for reference tracking.

\textbf{(2) M1-b:} No penalty ($\varphi=0$) for reference tracking.

\textbf{(3) M1-c:} Without reference for opportunity cost.

\begin{figure}[!t] 
    \setlength{\abovecaptionskip}{-0.1cm}  
    \setlength{\belowcaptionskip}{-0.1cm}   
    \centerline{\includegraphics[width=0.95\columnwidth]{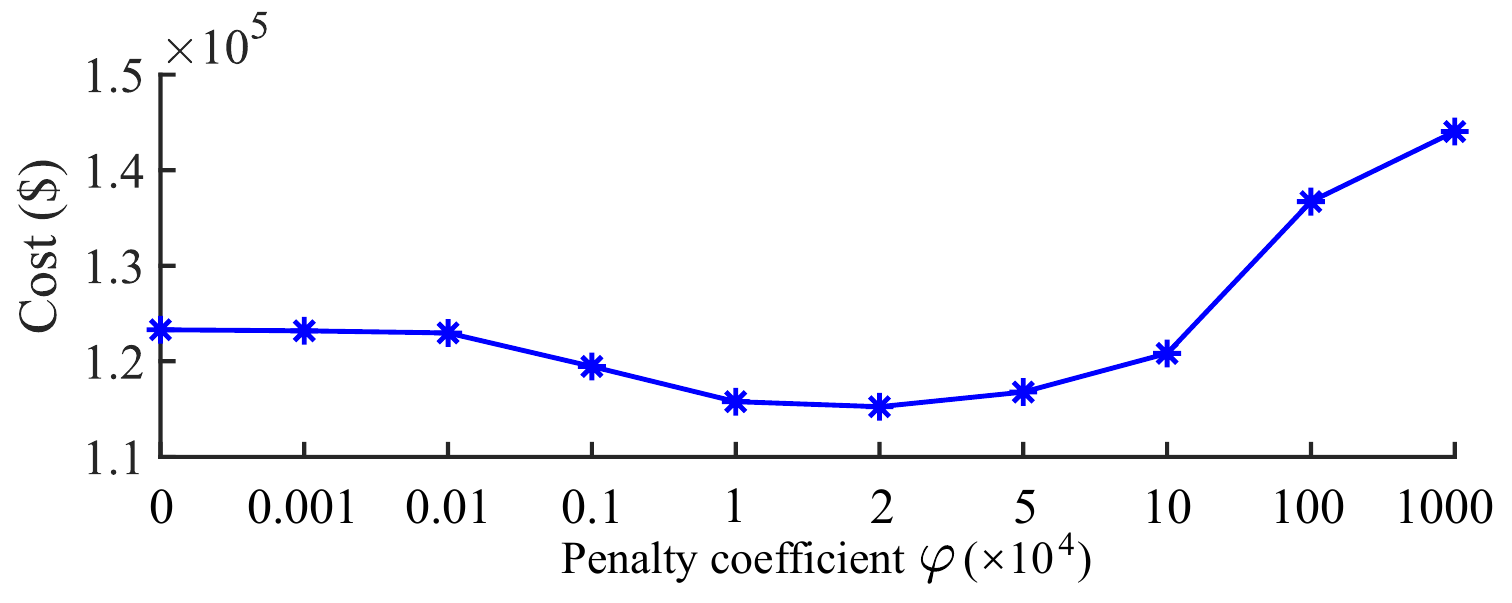}}
    \caption{Cost of M1 with different penalty coefficients for reference tracking.}
    \label{weight_of_SOC}
\end{figure}

\begin{figure}[!] 
    \setlength{\abovecaptionskip}{-0.1cm}  
    \setlength{\belowcaptionskip}{-0.1cm}   
\centerline{\includegraphics[width=0.95\columnwidth]{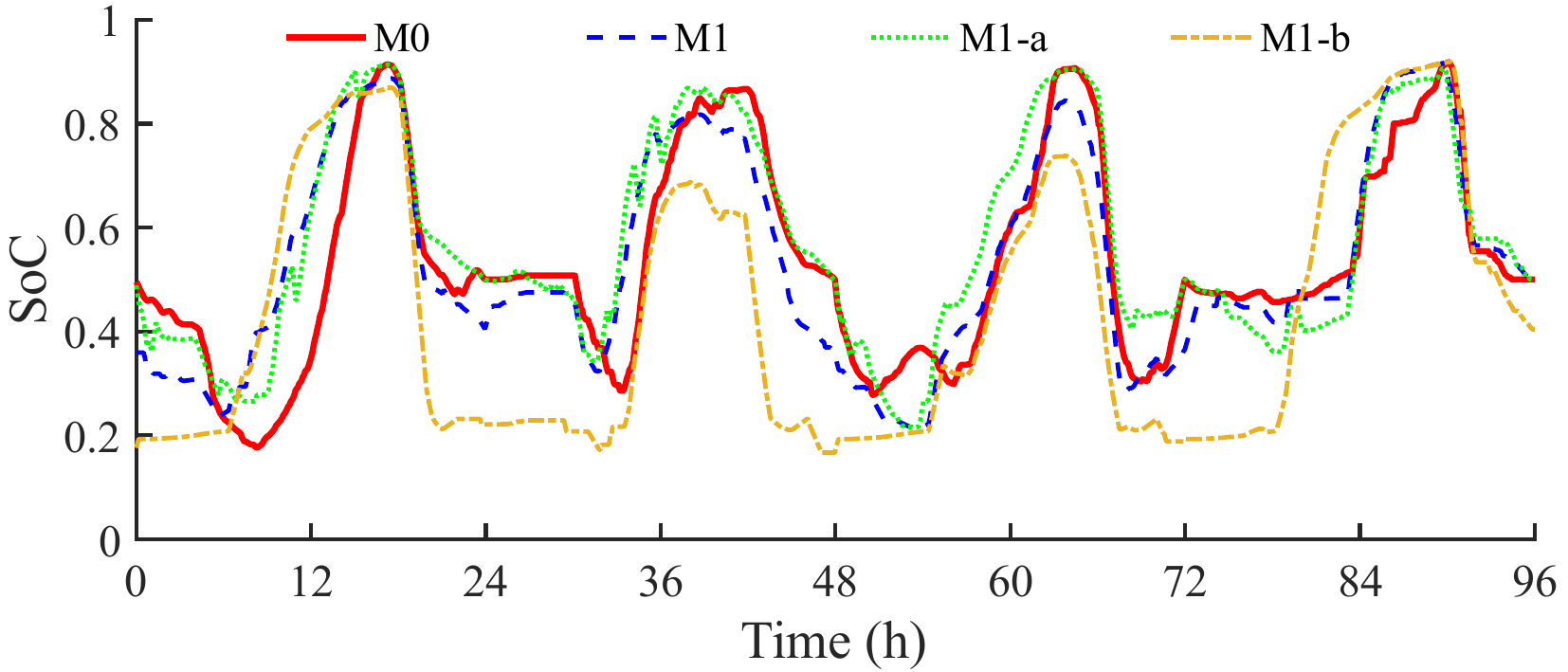}}
    \caption{Comparison of SoC among M0, M1, M1-a and M1-b for day 5-8.}
    \label{SOC comparison} 
\end{figure}

\begin{figure}[!] 
    \setlength{\abovecaptionskip}{-0.1cm}  
    \setlength{\belowcaptionskip}{-0.1cm}   
    \centerline{\includegraphics[width=0.95\columnwidth]{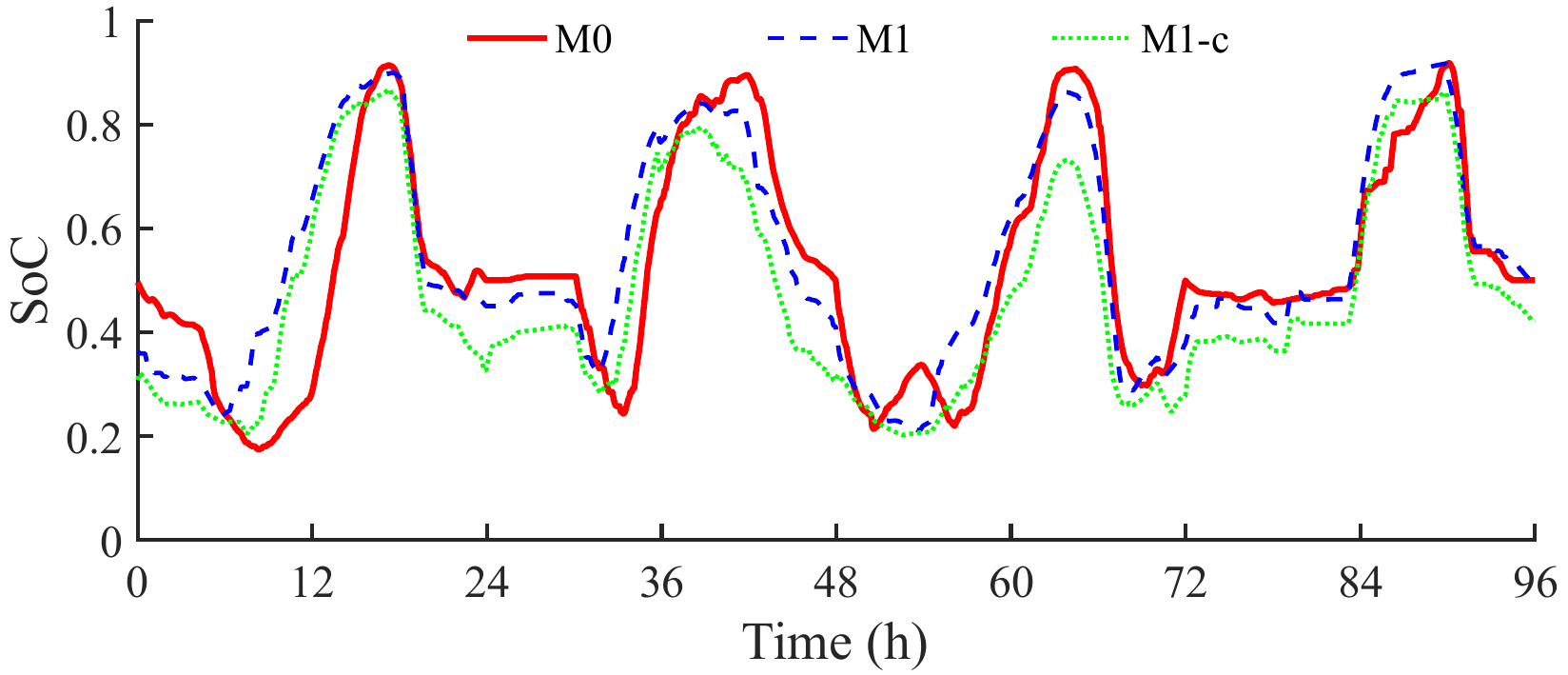}}
    \caption{Comparison of SoC among M0, M1 and M1-c for day 5-8.}
    \label{tie-line power comparison}
\end{figure}

We summarize the results in Table~\ref{different steps}. Due to the non-anticipativity of netload and prices, references do not always approximate the optimal baseline. For instance, during sudden price spikes (hours 67–71 in Figure~\ref{SOC comparison}), immediate GES discharge is preferable to strictly following the reference, which explains the higher costs of M1-a. Conversely, without a reference, the dispatch decisions (M1-b) become excessively aggressive, causing insufficient reserves for future operations. For example, during hours 41–43, SoC quickly dropped to its minimum, leaving inadequate discharge capacity when prices unexpectedly rose during hours 47–51. Therefore, appropriately tracking the reference is essential.

Additionally, as shown in Figure~\ref{tie-line power comparison}, the SoC in M1-c remains consistently lower than that in M1. This occurs because, without considering the opportunity cost, GES tends to discharge according to the reference whenever prices exceed physical discharge costs, thus behaving myopically. In contrast, incorporating opportunity costs effectively captures average future price trends, enabling GES to strategically perform peak-valley arbitrage. By appropriately sacrificing immediate benefits, this strategy ultimately achieves lower operational costs in the long run. This highlights the importance of incorporating opportunity costs in volatile, grid-aware operating environments.

{\color{blue}\begin{figure}[!t] 
    \setlength{\abovecaptionskip}{-0.1cm}  
    \setlength{\belowcaptionskip}{-0.1cm}   
    \centerline{\includegraphics[width=0.95\columnwidth]{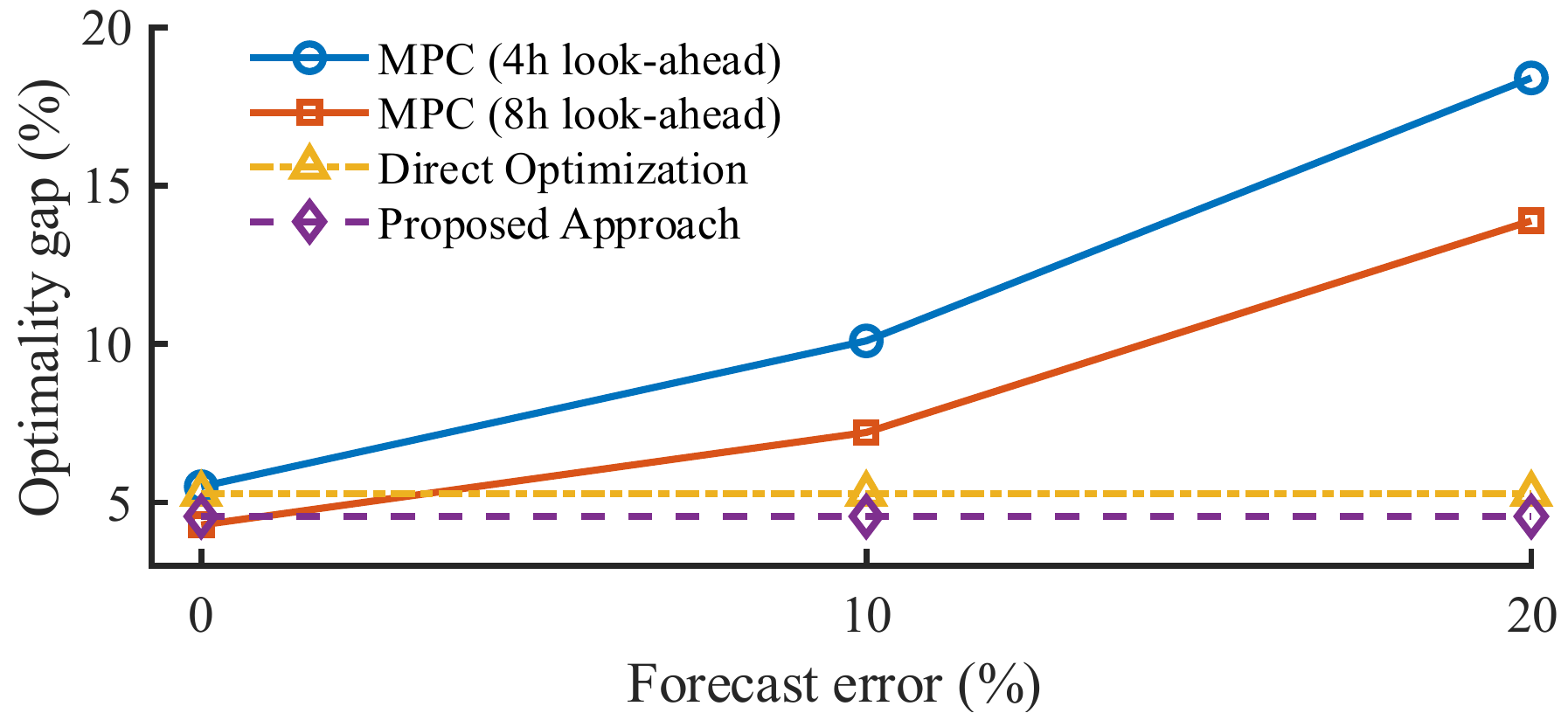}}
    \caption{{\color{blue}Optimality gap of different methods under varying forecast errors and look-ahead window lengths.}}
    \label{large_comparison}
\end{figure}}

{\color{blue}
\subsubsection{Result sensitivity to forecast error and look-ahead window length}
To further examine the robustness of the OCO approach, we analyze its sensitivity under varying forecast errors and look-ahead window lengths, while also providing a comparative analysis with MPC. The optimality gaps, defined relative to M0, which solves the OED problem in~\eqref{OED} with perfect knowledge of uncertainties, are summarized in Figure~\ref{large_comparison}. It is observed that the proposed OCO method achieves the smallest optimality gap and demonstrates remarkable robustness, remaining unaffected by varying forecast errors and look-ahead window lengths. In contrast, the MPC method exhibits significant sensitivity. Its optimality gap increases notably with higher forecast errors and shorter horizons. For instance, under a 4-hour horizon, the gap increases from 5.50\% (0\% error) to 18.40\% (20\% error). Although extending the look-ahead window to 8 hours can partially mitigate this issue—reducing the gap range from 4.29\% to 13.89\%—obtaining accurate long-horizon forecasts is practically challenging, particularly for small-scale microgrids with volatile renewable generation. Thus, the MPC's reliance on forecast accuracy represents a critical limitation compared to the proposed prediction-free OCO method.

We further investigated a scenario where the dispatch problem~\eqref{tracking} is solved directly at each step using only currently revealed uncertainties, without employing the OCO framework. A two-month case study under this setting results in an optimality gap of 5.27\% and a voltage satisfaction rate of 95.79\%. By comparison, the proposed OCO algorithm achieves a lower optimality gap of 4.56\% and an improved satisfaction rate of 98.62\%. These results clearly demonstrate the advantages of the OCO approach, which effectively addresses non-anticipativity constraints by approximating the evolution of uncertainties through a feedback-based penalty mechanism. However, since uncertainties remain unknown prior to decision-making in the OCO framework, it cannot fully capture future uncertainties, resulting in a voltage satisfaction rate below 100\%.
}
\rev{\subsubsection{Computational efficiency and scalability} To evaluate the computational efficiency and scalability of the proposed approach on large-scale networks, we conduct additional case studies on systems of different sizes, including the IEEE 69-bus and the IEEE 141-bus test systems~\cite{das2008optimal,khodr2008maximum}, and both the number of buses and the number of GES units are varied. The \emph{Average CPU Time} represents the average computational time required to solve each dispatch interval, and the \emph{Optimality Gap} is evaluated relative to the offline optimal solution. The results in Table~\ref{tab:scalability} show that as the system scale increases, the average CPU time grows slowly and approximately linearly with the number of buses and GES units. Additionally, the optimality gap slightly decreases as more GES units are included. Even in the largest tested scenario (141 buses with 64 GES units), the average computation time per 5-minute dispatch interval is only 1.58s, comfortably meeting real-time requirements. Moreover, the optimality gap consistently remains below 4.6\%, and the voltage satisfaction rate exceeds 98.5\% in all cases. These results confirm that the proposed approach has excellent computational efficiency and scalability, making it suitable for direct implementation in larger networks without compromising performance.}
\begin{table}[h]
\centering
\rev{\caption{Computational efficiency and scalability performance of the proposed approach}\label{tab:scalability}}
\renewcommand{\arraystretch}{1.1}
\setlength{\tabcolsep}{4pt}
\begin{tabular}{ccccc}
\toprule
\makecell{Number \\ of Bus} & \makecell{Number \\ of GES} & \makecell{Average CPU \\ Time (s)} & \makecell{Optimality \\ Gap (\%)} & \makecell{Voltage Satisfaction\\  Rate (\%)} \\
\midrule
33  & 16 & 0.49 & 4.56 & 98.62 \\
33  & 32 & 0.57 & 4.55 & 98.68 \\
69  & 16 & 0.68 & 4.53 & 98.57 \\
69  & 32 & 0.76 & 4.54 & 98.63 \\
69  & 48 & 0.91 & 4.54 & 98.66 \\
141 & 16 & 1.00 & 4.51 & 98.66 \\
141 & 32 & 1.15 & 4.52 & 98.68 \\
141 & 48 & 1.38 & 4.51 & 98.70 \\
141 & 64 & 1.58 & 4.51 & 98.72 \\
\bottomrule
\end{tabular}
\end{table}

\section{Conclusion}\label{Conclusion}

This paper proposes a novel prediction-free two-stage coordinated dispatch framework for the real-time dispatch of grid-connected microgrid with GES. We generate the hindsight SoC trajectories of GES offline with historical scenarios. Subsequently, we synthesize and dynamically update online references for both SoC and opportunity cost via kernel regression and leverage historical knowledge. We propose an \rev{adaptive Lagrange multiplier-based} OCO algorithm with reference tracking for global vision and expert-tracking for step-size updates. Numerical studies using ground-truth data from the Australian Energy Market Operator demonstrate that even with the identical reference tracking, the proposed method outperforms MPC and Lyapunov optimization methods, achieving operational cost reductions of 5.0\% and 6.2\%, and reducing voltage violations by 9.1\% and 0.8\%, respectively. Moreover, reference tracking substantially mitigates myopic nature of OCO, with the references for SoC and opportunity cost contributing approximately 6.5\% and 3.9\% to cost reductions, respectively. And dynamically updated references better capture real-time uncertainties compared to static day-ahead references, reducing operational costs by approximately 3.0\%. \rev{Sensitivity analysis confirms the robustness, computational efficiency, and scalability of the proposed method.} With rigorous theoretical guarantees and a prediction-free and explainable learning design, the proposed framework provides microgrid operators with a near-optimal, reliable, and adaptable real-time dispatch tool.

We discuss the potential extension of the proposed framework to more practical microgrid operations involving nonconvexities and abnormal states. In future work, we will continue exploring these aspects and investigate the integration of the OCO algorithm with distributed optimization techniques.
\appendix

\subsection{Proof of the relaxation of complementarity constraint} \label{appendix A}

We prove that simultaneous charging and discharging do not occur for any GES unit $s$ at time $t$, i.e., $P_{s\text{,}t}^{+}P_{s\text{,}t}^{-} = 0$.

We prove this by contradiction. Suppose there exists an optimal solution where $P_{s\text{,}t}^{+} > 0\text{, } P_{s\text{,}t}^{-} > 0.$ By substituting the power balance constraints~\eqref{P_flow} into the objective function~\eqref{objective}, we obtain the operational cost associated with GES unit $s$ at time $t$ as follows:
\begin{equation}\label{proof_102}
C_{s\text{,}t}^{\mathrm{GES}}= ({c}_{s}^{+}-c_{g\text{,}t})P_{s\text{,}t}^{+}+({c}_{s}^{-}+c_{g\text{,}t})P_{s\text{,}t}^{-}
\end{equation}

We now analyze two cases as follows:

Case 1: $P_{s\text{,}t}^{-}>P_{s\text{,}t}^{+}>0$.

We can find an alternative solution that $\tilde{P}_{s\text{,}t}^{-}={P}_{s\text{,}t}^{-}-{P}_{s\text{,}t}^{+}>0 \text{, }\tilde{P}_{s\text{,}t}^{+}=0.$ Then, we compute the cost difference between this alternative solution and the optimal solution as follows:
\begin{equation}\label{proof_104}
\tilde{C}_{s\text{,}t}^{\mathrm{GES}} - C_{s\text{,}t}^{\mathrm{GES}}=-{c}_{s}^{+}{P}_{s\text{,}t}^{+}<0
\end{equation}

According to~\eqref{GES_SOC}, the corresponding SoC difference is:

\begin{equation}
\tilde{E}_{s\text{,}t}-E_{s\text{,}t}={P}_{s\text{,}t}^{+}\left(\frac{1}{{\eta}_{s\text{,}t}}-{\eta}_{s\text{,}t}\right)\geq0
\end{equation}

This indicates that the alternative solution improves both cost and SoC, which contradicts the assumption of optimality.

Case2: $P_{s\text{,}t}^{+}>P_{s\text{,}t}^{-}>0$.

We can find an alternative solution that $\tilde{P}_{s\text{,}t}^{+}={P}_{s\text{,}t}^{+}-{P}_{s\text{,}t}^{-}>0 \text{, }\tilde{P}_{s\text{,}t}^{-}=0.$ Then, we compute the cost difference between this alternative solution and the optimal solution as follows:
\begin{equation}\label{proof_107}
\tilde{C}_{s\text{,}t}^{\mathrm{GES}} - C_{s\text{,}t}^{\mathrm{GES}}=-{c}_{s}^{-}{P}_{s\text{,}t}^{-}<0
\end{equation}

According to~\eqref{GES_SOC}, the corresponding SoC difference is:

\begin{equation}
\tilde{E}_{s\text{,}t}-E_{s\text{,}t}={P}_{s\text{,}t}^{-}\left(\frac{1}{{\eta}_{s\text{,}t}}-{\eta}_{s\text{,}t}\right)\geq0
\end{equation}

This indicates that the alternative solution improves both cost and SoC, which contradicts the assumption of optimality.

Since both cases yield contradictions, our initial assumption of simultaneous charging and discharging does not hold. Hence, we have completed the proof.

\subsection{Proof of the bound on dynamic regret} \label{appendix B}
Let $\{x_{i,t}\}(\forall i\in [N])$ and $\{x_t\}$ be the sequences generated by Algorithm 1, and let $\{y_t\}$ be an arbitrary sequence in $X$. Given the convexity of $f_t$ and Assumption 3, we have:
\begin{align}\label{proof3}
&f_t(x_{i\text{,}t})-f_t(y_t)\leq\left\langle\partial f_t(x_{i\text{,}t})\text{, } x_{i\text{,}t}-y_t\right\rangle\\
&=\left\langle\partial f_t(x_{i\text{,}t})\text{, } x_{i\text{,}t}-x_{i\text{,}t+1}\right\rangle+\left\langle\partial f_t(x_{i\text{,}t})\text{, } x_{i\text{,}t+1}-y_t\right\rangle\notag\\
&\leq J\|x_{i\text{,}t}-x_{i\text{,}t+1}\|+\left\langle\partial f_t(x_{i\text{,}t})\text{, } x_{i\text{,}t+1}-y_t\right\rangle\notag\\
&\leq \frac{J^2\alpha_{i\text{,}t}}{2}+\frac{1}{2\alpha_{i\text{,}t}}\|x_{i\text{,}t}-x_{i\text{,}t+1}\|^2+\left\langle\partial f_t(x_{i\text{,}t})\text{, } x_{i\text{,}t+1}-y_t\right\rangle\notag
\end{align}

The first two terms of the last inequality are derived from the Arithmetic-Geometric Mean inequality. For the third term of~\eqref{proof3}, we have:
\begin{align}\label{proof4}
&\left\langle\partial f_t(x_{i\text{,}t})\text{, } x_{i\text{,}t+1}-y_t\right\rangle 
\\&=\left\langle\beta_{t}(\nu_{i\text{,}t}^{\top}\partial[h_t(x_{i\text{,}t+1})]_+)\text{, } y_t-x_{i\text{,}t+1}\right\rangle\notag 
\\&+\left\langle\partial f_t(x_{i\text{,}t})+\beta_{t}(\nu_{i\text{,}t}^{\top}\partial[h_t(x_{i\text{,}t+1})]_+)\text{, } x_{i\text{,}t+1}-y_t\right\rangle\notag
\end{align}

Since $g_t$ is convex, it is straightforward to verify that $[g_t(\cdot)]_+$ is also convex. Furthermore, from~\eqref{Q_update} and the parameter settings in~\eqref{parameter}, we have $\nu_{i,t}\geq0$. Therefore, by the convexity of $[h_t(\cdot)]_+$, we have:
\begin{align}\label{proof5}
&\left\langle\beta_{t}(\nu_{i\text{,}t}^{\top}\partial[h_t(x_{i\text{,}t+1})]_+)\text{,}\ y_t-x_{i\text{,}t+1}\right\rangle\\
&=\beta_{t}\left\langle \nu_{i\text{,}t}\text{,}\ \partial[h_t(x_{i\text{,}t+1})]_+\text{,}\ y_t-x_{i\text{,}t+1}\right\rangle\notag\\
&\leq\beta_{t}\left\langle \nu_{i,t},\ [h_t(y_t)]_+-[h_t(x_{i\text{,}t+1})]_+\right\rangle\notag\\
&=\beta_{t}\left\langle \nu_{i\text{,}t}\text{,}\ [h_t(y_t)]_+\right\rangle-\beta_{t}\left\langle \nu_{i\text{,}t}\text{,}[h_t(x_{i\text{,}t+1})]_+\right\rangle\notag
\end{align}

Note that~\eqref{x_update} can be equivalently rewritten as the following regularized optimization form:
\begin{equation}\label{proof6}
x_{i,t+1}=\arg\min_{x\in X}\{\hat{f}_{t}(x)+\|x - x_{i\text{,}t}\|^2\}
\end{equation}
where $\hat{f}_{t}(x)=\alpha_{i\text{,}t}\left\langle\partial f_{t}(x_{i\text{,}t})\text{,}\ x-x_{i\text{,}t}\right\rangle+\alpha_{i\text{,}t}\beta_{t}\left\langle \nu_{i\text{,}t},\text{,}\ [h_{t}(x)]_+\right\rangle$

It is straightforward to verify that $\partial \hat{f}_{t}$ exists and is bounded due to the bounded gradient assumption.

Applying Lemma 1 from reference~\cite{yi2020distributed} to~\eqref{proof6}, we have:
\begin{align}\label{proof7}
&\left\langle\partial f_t(x_{i\text{,}t})+\beta_{t}(\nu_{i\text{,}t}^\top\partial[h_t(x_{i\text{,}t+1})]_+)\text{,}\ x_{i\text{,}t+1}-y_t\right\rangle\\
&=\frac{1}{\alpha_{i,t}}\left\langle x_{i\text{,}t+1}-y_t\text{,}\ \partial \hat{f}_{t}(x_{i\text{,}t+1})\right\rangle\notag\\
&\leq\frac{1}{\alpha_{i\text{,}t}}\left(\|y_t-x_{i\text{,}t}\|^2-\|y_t-x_{i\text{,}t+1}\|^2-\|x_{i\text{,}t+1}-x_{i\text{,}t}\|^2\right)\notag
\end{align}

Combining~\eqref{proof3}-\eqref{proof5}, and~\eqref{proof7}, we have:
\begin{align}\label{proof8}
&f_t(x_{i\text{,}t})-f_t(y_t)\\
&\leq \frac{J^2\alpha_{i\text{,}t}}{2}+\frac{1}{2\alpha_{i\text{,}t}}\|x_{i\text{,}t}-x_{i\text{,}t+1}\|^2\notag\\
&\quad+\beta_{t}\left\langle \nu_{i\text{,}t}\text{,}[h_t(y_t)]_+\right\rangle - \beta_{t}\left\langle \nu_{i\text{,}t}\text{,}[h_t(x_{i,t+1})]_+\right\rangle\notag\\
&\quad+\frac{1}{\alpha_{i\text{,}t}}\left(\|y_t-x_{i\text{,}t}\|^2-\|y_t-x_{i\text{,}t+1}\|^2-\|x_{i\text{,}t+1}-x_{i\text{,}t}\|^2\right)\notag\\
&=\frac{J^2\alpha_{i\text{,}t}}{2}+\frac{1}{\alpha_{i\text{,}t}}\left(\|y_t-x_{i\text{,}t}\|^2-\|y_t-x_{i\text{,}t+1}\|^2\right)\notag\\
&\quad+\beta_{t}\left\langle \nu_{i\text{,}t}\text{,}[h_t(y_t)]_+\right\rangle\notag\\
&\quad-\frac{1}{2\alpha_{i,t}}\|x_{i\text{,}t+1}-x_{i\text{,}t}\|^2-\beta_{t}\left\langle \nu_{i\text{,}t}\text{,}[h_t(x_{i\text{,}t+1})]_+\right\rangle\notag\notag\\
&\leq\frac{J^2\alpha_{i\text{,}t}}{2}+\frac{1}{\alpha_{i\text{,}t}}\left(\|y_t-x_{i\text{,}t}\|^2-\|y_t-x_{i\text{,}t+1}\|^2\right)\notag\\
&\quad+\beta_{t}\left\langle \nu_{i\text{,}t}\text{,}[h_t(y_t)]_+\right\rangle\notag
\end{align}

By substituting $y_t$ with with the optimal solution $x_t^*$ in~\eqref{proof8}, we have: 
\begin{align}\label{proof9}
&\sum_{t=1}^Tf_t(x_{i\text{,}t})-\sum_{t=1}^Tf_t(x_t^*)\\
&\leq\sum_{t=1}^T\frac1{\alpha_{i\text{,}t}}(\parallel x_t^*-x_{i\text{,}t}\parallel^2-\parallel x_t^*-x_{i\text{,}t+1}\parallel^2)+\sum_{t=1}^T\frac{J^2\alpha_{i\text{,}t}}2\notag
\end{align}

Using the boundedness assumption in~\eqref{assump1}, the setting of $\alpha_{i,t}$ in~\eqref{parameter}, and the definition of the path-length $P_x$, we have:
\begin{align}\label{proof10}
&\sum_{t=1}^{T}\frac{1}{\alpha_{i\text{,}t}}\left(\|x_t^*-x_{i\text{,}t}\|^2-\|x_t^*-x_{i\text{,}t+1}\|^2\right)\\
=&\frac{1}{2^{i-1}}\sum_{t=1}^{T}\left(t^{\frac{1+2\chi}{2}}\|x_t^*-x_{i\text{,}t}\|^2 - (t+1)^{\frac{1}{2}+\chi}\|x_{t+1}^*-x_{i\text{,}t+1}\|^2\right)+\notag\\
&\frac{1}{2^{i-1}}\sum_{t=1}^{T}\left((t+1)^{\frac{1+2\chi}{2}}\|x_{t+1}^*-x_{i\text{,}t+1}\|^2 - t^{\frac{1+2\chi}{2}}\|x_{t+1}^*-x_{i\text{,}t+1}\|^2\right)\notag\\
&+\frac{1}{2^{i-1}}\sum_{t=1}^{T}\left(t^{\frac{1+2\chi}{2}}\|x_{t+1}^*-x_{i\text{,}t+1}\|^2 - t^{\frac{1+2\chi}{2}}\|x_t^*-x_{i\text{,}t+1}\|^2\right)\notag\\
\leq&\frac{1}{2^{i-1}}\|x_1^*-x_{i\text{,}1}\|^2 + \frac{1}{2^{i-1}}\sum_{t=1}^{T}\left((t+1)^{\frac{1+2\chi}{2}}-t^{\frac{1+2\chi}{2}}\right)W^2\notag\\
&+\frac{1}{2^{i-1}}\sum_{t=1}^{T}t^{\frac{1+2\chi}{2}}\left\langle x_{t+1}^*-x_t^*\text{,}x_{t+1}^*-x_{i\text{,}t+1}+x_t^*-x_{i\text{,}t+1}\right\rangle\notag\\
\leq&\frac{1}{2^{i-1}}W^2+\frac{1}{2^{i-1}}\left((T+1)^{\frac{1+2\chi}{2}}-1\right)W^2+\notag\\
&\frac{1}{2^{i-1}}\sum_{t=1}^{T}t^{\frac{1+2\chi}{2}}\|x_{t+1}^*-x_t^*\|\left(\|x_{t+1}^*-x_{i\text{,}t+1}\|+\|x_t^*-x_{i\text{,}t+1}\|\right)\notag\\
\leq&\frac{1}{2^{i-1}}\left(W^2(T+1)^{\frac{1+2\chi}{2}}+2WP_xT^{\frac{1+2\chi}{2}}\right)\notag
\end{align}

For the second term of~\eqref{proof9}, we have:
\begin{align}\label{proof11}
\sum_{t=1}^T\frac{J^2\alpha_{i\text{,}t}}{2}&\leq\frac{2^{i-1}J^2}{2}\sum_{t=1}^T\frac{1}{t^{\frac{1+2\chi}{2}}}\\
&\leq\frac{2^{i-1}J^2}{2}\left(\int_1^T t^{-(\frac{1+2\chi}{2})}\ \text{,}dt+1\right)\notag\\
&=\frac{2^{i-1}J^2}{2}\left(\frac{2}{1-2\chi}T^{\frac{1-2\chi}{2}}-\frac{1+2\chi}{1-2\chi}\right)\notag\\
&\leq\frac{2^{i-1}J^2}{1-2\chi}T^{\frac{1-2\chi}{2}}\notag
\end{align}

Combining~\eqref{proof9}-\eqref{proof11}, we have:
\begin{align}\label{proof12}
\sum_{t=1}^Tf_t(x_{i\text{,}t})-\sum_{t=1}^Tf_t(x_t^*)&\leq\frac2{2^{i-1}}(W^2T^{\frac{1+2\chi}{2}}(1+\frac{P_x}{W})
\\&+\frac{2^{i-1}G^2}{1-2\chi}T^{\frac{1-2\chi}{2}}\notag
\end{align}

Let $i_0=[\frac12\log_2(1+\frac{P_x}{W})]+1\in[N]$ such that:
\begin{equation}\label{proof13}
\begin{aligned}
2^{i_0-1}\leq\sqrt{1+\frac{P_x}{W}}\leq2^{i_0}
\end{aligned}
\end{equation}
Substitute $i_{0}$ in~\eqref{proof12}, and combining~\eqref{proof12} and~\eqref{proof13} yields:
\begin{equation}\label{proof14}
\begin{aligned}
\sum_{t=1}^Tf_t(x_{i_0\text{,}t})-\sum_{t=1}^Tf_t(x_t^*)\leq(4W^2+\frac{J^2}{1-2\chi})T^{\frac{1+2\chi}{2}}\sqrt{1+\frac{P_x}{W}}
\end{aligned}
\end{equation}
From~\eqref{final_decision} and the convexity of $f_{t}$, we have:
\begin{align}\label{proof15}
f_t(x_t)-f_t(x_{i_0\text{,}t})&\leq\left\langle \partial f_t(x_t)\text{, } x_t - x_{i_0\text{,}t}\right\rangle\\
&= \ell_t(x_t)-\ell_t(x_{i_0\text{,}t})\notag
\end{align}

Applying Lemma 3 in reference~\cite{yi2021regret} to~\eqref{loss} and~\eqref{final_decision} yields:
\begin{align}\label{proof16}
\sum_{t=1}^T\ell_t(x_t)-\min_{i\in[N]}\{\sum_{t=1}^T\ell_t(x_{i\text{,}t})+\frac1\gamma\ln\frac1{\rho_{i\text{,}1}}\}\leq\frac{\gamma(JW)^2T}2
\end{align}

Combining~\eqref{proof15}-\eqref{proof16}, and the definition of $\gamma$ in~\eqref{parameter} yields:
\begin{align}\label{proof17}
\sum_{t=1}^Tf_t(x_t)-\sum_{t=1}^Tf_t(x_{i_0\text{,}t})&\leq \sum_{t=1}^T \ell_t(x_t)-\sum_{t=1}^T \ell_t(x_{i_0\text{,}t})\\
&\leq \frac{\gamma(JW)^2T}{2}+\frac{1}{\gamma}\ln\frac{1}{\rho_{i_0\text{,}1}}\notag\\
&=\frac{(JW)^2\sqrt{T}}{2}+\sqrt{T}\ln\frac{1}{\rho_{i_0\text{,}1}}\notag
\end{align}

From the initialization step of Algorithm~\ref{alg1}, we have:
\begin{align}\label{proof18}
\ln\frac1{\rho_{i_0\text{,}1}}\leq\ln(i_0(i_0+1))&\leq2\ln(i_0+1)\\&\leq2\ln(\left\lfloor\frac12\log_2(1+\frac{P_x}{W})\right\rfloor+2)\notag
\end{align}

Combining~\eqref{proof14},~\eqref{proof17} and~\eqref{proof18} yields dynamic regret bound:
\begin{align}\label{result_regret}
\text{Regret}&=\sum_{t=1}^{T}f_t(x_t)-\sum_{t=1}^{T}f_t(x_t^*)\\
&=\sum_{t=1}^{T}f_t(x_t)-\sum_{t=1}^{T}f_t(x_{i_0\text{,}t})+\sum_{t=1}^{T}f_t(x_{i_0\text{,}t})-\sum_{t=1}^{T}f_t(x_t^*)\notag\\
&\leq\frac{(JW)^2\sqrt{T}}{2}+\sqrt{T}\ln\frac{1}{\rho_{i_0\text{,}1}}\notag\\
&\quad+(4W^2+\frac{G^2}{1-2\chi})T^{\frac{1+2\chi}{2}}\sqrt{1+\frac{P_x}{W}}\notag\\
&\leq\frac{(JW)^2\sqrt{T}}{2}+2\sqrt{T}\ln\left([\frac{1}{2}\log_2\left(1+\frac{P_x}{W}\right)]+2\right)\notag\\
&\quad+(4W^2+\frac{J^2}{1-2\chi})T^{\frac{1+2\chi}{2}}\sqrt{1+\frac{P_x}{W}}\notag\\
&=O\left(T^{\frac{1+2\chi}{2}}\sqrt{1+P_x}\right)\notag
\end{align}
Hence, we have finished the proof.

\subsection{Proof of the bound on constraint violations} \label{appendix C}
We first define $\tilde{f}_t(x)=\alpha_{i\text{,}t}\left\langle\partial f_t(x_{i\text{,}t})\text{,}\ x-x_{i\text{,}t}\right\rangle+\alpha_{i\text{,}t}\beta_t\left\langle \nu_{i\text{,}t}\text{,}\ [h_t(x)]_+\right\rangle+\parallel x-x_{i\text{,}t}\parallel^2$. Note that $\tilde{f}_t$ and $\left|\left|x-x_t\right|\right|^2$ are 2-strongly convex, we have:
\begin{equation}\label{proof_2}
\begin{aligned}
\tilde{f}_{t}(x)\geq \tilde{f}_{t}(y)+\left\langle\partial \tilde{f}_{t}(y)\text{,}\ x-y\right\rangle+\parallel x-y\parallel^2
\end{aligned}
\end{equation}

According to~\eqref{x_update}, we have $x_{i,t+1}=\arg\min_{x\in X}\tilde{f}(x)$. Based on the first-order optimality condition, we have:
\begin{equation}\label{proof_3}
\begin{aligned}
\left\langle\partial \tilde{f}_{t}(x_{i\text{,}t+1})\text{,}\ x-x_{i\text{,}t+1}\right\rangle \geq0
\end{aligned}
\end{equation}

By substituting $y=x_{i,t+1}$ and $x=x_t^*$ in~\eqref{proof_2}, and combining~\eqref{proof_3} yields:
\begin{align}\label{proof_4}
&\alpha_{i\text{,}t}\left\langle\partial f_t(x_{i\text{,}t})\text{,}\ x_{i\text{,}t+1}-x_{i\text{,}t}\right\rangle\\
&+\parallel x_{i\text{,}t+1}-x_{i\text{,}t}\parallel^2+\alpha_{i\text{,}t}\beta_t\left\langle \nu_{i\text{,}t}\text{,}\ [h_t(x_{i\text{,}t+1})]_+\right\rangle\notag\\
&\leq \tilde{f}_{t}(x_t^*) - \left\langle \partial \tilde{f}_{t}(x_{i\text{,}t+1})\text{,} x_t^*-x_{i\text{,}t+1}\right\rangle - \|x_t^*-x_{i\text{,}t+1}\|^2\notag\\
&\leq \alpha_{i\text{,}t}\left\langle\partial f_t(x_{i\text{,}t})\text{,} x_t^*-x_{i\text{,}t}\right\rangle+\alpha_{i\text{,}t}\beta_t\left\langle \nu_{i\text{,}t} [h_t(x_t^*)]_+\right\rangle\notag\\
&\quad+\|x_t^*-x_{i\text{,}t}\|^2 - \|x_t^*-x_{i\text{,}t+1}\|^2\notag
\end{align}

\noindent where the second inequality holds due to the optimality condition~\eqref{proof_3} and the definition of $\tilde{f}_{t}$.

We add $\alpha_{i\text{,}t}f_t(x_{i\text{,}t})$ to both sides and we have:
\begin{align}\label{proof_5}
&\alpha_{i\text{,}t}f_t(x_{i\text{,}t})-\alpha_{i\text{,}t}(f_t(x_{i\text{,}t})+<\partial f_t(x_{i\text{,}t})\text{,}\ x_t^*-x_{i\text{,}t}>)\\
&+\alpha_{i\text{,}t}\beta_t\left\langle \nu_{i\text{,}t}\text{,}\ [h_t(x_{i\text{,}t+1})]_+\right\rangle\notag
\\
&\leq\alpha_{i\text{,}t}f_t(x_{i\text{,}t})-\alpha_{i\text{,}t}f_t(x_t^*)+\alpha_{i\text{,}t}\beta_t\left\langle \nu_{i\text{,}t}\text{,}\ [h_t(x_{i\text{,}t+1})]_+\right\rangle\notag
\\&\leq\alpha_{i\text{,}t}\left\langle\partial f_t(x_{i\text{,}t})\text{,}\ x_{i\text{,}t}-x_{i\text{,}t+1}\right\rangle+\alpha_{i\text{,}t}\beta_t\left\langle \nu_{i\text{,}t}\text{,}\ [h_t(x_t^*)]_+\right\rangle\notag\\
&\quad-\parallel x_{i\text{,}t+1}-x_{i\text{,}t}\parallel^2+\parallel x_t^*-x_{i\text{,}t}\parallel^2-\parallel x_t^*-x_{i\text{,}t+1}\parallel^2\notag
\end{align}

We have $[h_t(x_t^*)]_+=0$, combining~\eqref{assump3} and~\eqref{proof_5} yields:
\begin{align}\label{proof_6}
&\alpha_{i\text{,}t}f_{t}(x_{i\text{,}t})-\alpha_{i\text{,}t}f_{t}(x_{t}^{*})+\alpha_{i\text{,}t}\beta_{t}\left\langle \nu_{i\text{,}t}\text{,} [h_{t}(x_{i\text{,}t+1})]_+\right\rangle\\
&\leq\alpha_{i\text{,}t}\left\langle\partial f_{t}(x_{i\text{,}t})\text{,} x_{i\text{,}t}-x_{i\text{,}t+1}\right\rangle - \|x_{i\text{,}t+1}-x_{i\text{,}t}\|^2 \notag\\
&\quad+ \|x_{t}^{*}-x_{i\text{,}t}\|^2 - \|x_{t}^{*}-x_{i\text{,}t+1}\|^2\notag\\
&\leq\alpha_{i\text{,}t}\left\langle\partial f_{t}(x_{i\text{,}t})\text{,} x_{i\text{,}t}-x_{i\text{,}t+1}\right\rangle - \|x_{i\text{,}t+1}-x_{i\text{,}t}\|^2 +\|x_{t}^{*}-x_{i\text{,}t}\|^2\notag\\
&\quad - \|x_{t}^{*}-x_{i\text{,}t+1}\|^2+ \frac{\alpha_{i\text{,}t}^2\|\partial f_{t}(x_{i\text{,}t})\|^2}{4}- \frac{\alpha_{i\text{,}t}^2\|\partial f_{t}(x_{i\text{,}t})\|^2}{4}\notag\\
&\leq -\left\|\frac{\alpha_{i\text{,}t}}{2}\partial f_{t}(x_{i\text{,}t}) - (x_{i\text{,}t+1}-x_{i\text{,}t})\right\|^2 + \frac{\alpha_{i\text{,}t}^2\|\partial f_{t}(x_{i\text{,}t})\|^2}{4}\notag\\
&\quad+\|x_{t}^{*}-x_{i\text{,}t}\|^2 - \|x_{t}^{*}-x_{i\text{,}t+1}\|^2\notag\\
&\leq \frac{\alpha_{i\text{,}t}^2\|\partial f_{t}(x_{i\text{,}t})\|^2}{4} + \|x_{t}^{*}-x_{i\text{,}t}\|^2 - \|x_{t}^{*}-x_{i\text{,}t+1}\|^2\notag\\
&\leq \frac{\alpha_{i\text{,}t}^2 J^2}{4} + \|x_{t}^{*}-x_{i\text{,}t}\|^2 - \|x_{t}^{*}-x_{i\text{,}t+1}\|^2\notag
\end{align}

From~\eqref{proof_6}, we have:
\begin{align}\label{proof_7}
&\alpha_{i\text{,}t}\beta_t\left\langle \nu_{i\text{,}t}\text{,}\ [h_t(x_{i\text{,}t+1})]_+\right\rangle\leq\frac{\alpha_{i\text{,}t}^2J^2}4\\&+\alpha_{i\text{,}t}\mid f_t(x_{i\text{,}t})-f_t(x_t^*)\mid+\parallel x_t^*-x_{i\text{,}t}\parallel^2-\parallel x_t^*-x_{i\text{,}t+1}\parallel^2 \notag
\end{align}

From~\eqref{Q_update}, we have:
\begin{equation}\label{proof_8}
\begin{aligned}
\alpha_{i\text{,}t}\beta_t\left\langle \nu_{i\text{,}t}\text{,}\ [h_t(x_{i\text{,}t+1})]_+\right\rangle\geq\alpha_{i\text{,}t}\beta_t\theta_{i\text{,}t}\parallel[h_t(x_{i\text{,}t+1})]_+\parallel_1
\end{aligned}
\end{equation}

Combining~\eqref{parameter},~\eqref{proof_7} and~\eqref{proof_8} yields:
\begin{align}\label{proof_9}
\|[h_t(x_{i\text{,}t+1})]_+\|_1 &\leq\frac{\alpha_{i\text{,}t}J^2}{4\beta_t\theta_{i\text{,}t}}+\frac{|f_t(x_{i\text{,}t})-f_t(x_t^*)|}{\beta_t\theta_{i\text{,}t}}\\
&\quad+\frac{\|x_t^*-x_{i\text{,}t}\|^2-\|x_t^*-x_{i\text{,}t+1}\|^2}{\alpha_{i\text{,}t}\beta_t\theta_{i\text{,}t}}\notag\\
&\leq\frac{J^2}{4t^{2+\delta+\chi}}+\frac{|f_t(x_{i\text{,}t})-f_t(x_t^*)|}{2^{i-1}t^{\frac{3}{2}+\delta}}\notag\\
&\quad+\frac{\|x_t^*-x_{i\text{,}t}\|^2-\|x_t^*-x_{i\text{,}t+1}\|^2}{4^{i-1}t^{1+\delta-\chi}}\notag
\end{align}

Under Assumptions 1-3, summing~\eqref{proof_9} over $t$ yields:
\begin{align}\label{proof_10}
&\sum_{t=1}^{T}\|[h(x_{i\text{,}t+1})]_{+}\|_{1}\\
&\leq\frac{J^2}{4}\left(1+\int_1^T\frac{1}{t^{2+\delta+\chi}}dt\right)+\frac{F}{2^{i-1}}\left(1+\int_1^T\frac{1}{t^{\frac{3}{2}+\delta}}dt\right)\notag\\
&\quad+\frac{W^2}{4^{i-1}}\left(1+\int_1^T\frac{1}{t^{1+\delta-\chi}}dt\right)\notag\\
&\leq\frac{J^2}{2}+\frac{3F}{2^{i-1}}+\left(1+\frac{1}{\delta-\chi}\right)\frac{W^2}{4^{i-1}}\notag
\end{align}

By applying the Cauchy–Schwarz inequality, Assumption 3, and letting $\zeta=T^{\chi/2}$, we have:
\begin{align}\label{proof_11}
&\|[h_t(x_{i\text{,}t})]_+\|_1 - \|[h_t(x_{i\text{,}t+1})]_+\|_1\\
&\leq \left\langle \partial [h_t(x_{i\text{,}t})]_+\text{, } x_{i\text{,}t}-x_{i\text{,}t+1}\right\rangle\notag\\
&\leq \|\partial [h_t(x_{i\text{,}t})]_+\|\cdot\|x_{i\text{,}t}-x_{i\text{,}t+1}\|\notag\\
&\leq J\|x_{i\text{,}t}-x_{i\text{,}t+1}\| - \left(\frac{J^2}{4\zeta}+\zeta\|x_{i\text{,}t}-x_{i\text{,}t+1}\|^2\right) + \frac{J^2}{4\zeta} \notag\\
&\quad+ \zeta\|x_{i\text{,}t}-x_{i\text{,}t+1}\|^2\notag\\
&\leq J\|x_{i\text{,}t}-x_{i\text{,}t+1}\| - 2\sqrt{\frac{J^2}{4\zeta}\cdot\zeta\|x_{i\text{,}t}-x_{i\text{,}t+1}\|^2} + \frac{J^2}{4\zeta}\notag\\
&\quad+ \zeta\|x_{i\text{,}t}-x_{i\text{,}t+1}\|^2\notag\\
&\leq \frac{J^2}{4\zeta} + \zeta\|x_{i\text{,}t}-x_{i\text{,}t+1}\|^2\notag
\end{align}

Note that $[h_t(x_{i\text{,}t+1})]_+ > 0$ and $[h_t(x_t^*)]_+ = 0$, rearranging the terms of~\eqref{proof_5} yields:
\begin{align}\label{proof_42}
&\|x_{i\text{,}t+1}-x_{i\text{,}t}\|^2 \leq \alpha_{i\text{,}t}\left\langle \partial f_t(x_{i\text{,}t})\text{,} x_{i\text{,}t}-x_{i\text{,}t+1}\right\rangle  \\
&+ \|x_t^*-x_{i\text{,}t}\|^2- \|x_t^*-x_{i\text{,}t+1}\|^2+\alpha_{i\text{,}t}\left(f_t(x_t^*)-f_t(x_{i\text{,}t})\right)\notag
\end{align}

Under Assumptions 1-3, summing~\eqref{proof_42} over $t$ yields:
\begin{align}\label{proof_sum_42}
&\sum_{t=1}^{T}\|x_{i\text{,}t+1}-x_{i\text{,}t}\|^2\\
&\leq (F+JW)\sum_{t=1}^{T}\alpha_{i\text{,}t} + \|x_1^*-x_{i\text{,}1}\|^2\notag\\
&\leq 2^{i-1}(F+JW)\left(1+\int_{1}^{T}\frac{1}{t^{\frac{1}{2}+\chi}}dt\right)+W^2\notag\\
&\leq 2^{i-1}(F+JW)\left(\frac{2}{1-2\chi}T^{\frac{1}{2}-\chi}-\frac{1+2\chi}{1-2\chi}\right)+W^2\notag
\end{align}

Summing~\eqref{proof_sum_42} over $t$ and combining~\eqref{proof_11} yields:
\begin{align}\label{proof_69}
&\sum_{t=1}^T\|[h_t(x_{i\text{,}t})]_+\|_1-\|[h_t(x_{i\text{,}t+1})]_+\|_1\\
&\leq\frac{J^2}4T^{1-\frac\chi2}+2^{i-1}(F+JW)\frac2{1-2\chi}T^{\frac12-\frac\chi2}+W^2T^{\frac\chi2}\notag
\end{align}

Also, it is straightforward to show:
\begin{equation}\label{proof_70}
\sum_{t=1}^{T}\|[h_{t}(x_{i\text{,}t})]_+\|\leq \sum_{t=1}^{T}\|[h_{t}(x_{i\text{,}t})]_+\|_{1}
\end{equation}

Combining~\eqref{proof_10},~\eqref{proof_69}, and~\eqref{proof_70} yields:
\begin{align}\label{proof_72}
&\sum_{t=1}^{T}\|[h_{t}(x_{i\text{,}t})]_+\|\\
&\leq\sum_{t=1}^{T}\left(\|[h_{t}(x_{i\text{,}t})]_+\|_{1}-\|[h_{t}(x_{i\text{,}t+1})]_+\|_{1}\right)+\sum_{t=1}^{T}\|[h(x_{i\text{,}t+1})]_+\|_{1}\notag\\
&\leq\frac{J^{2}}{2}+\frac{3F}{2^{i-1}}+\left(1+\frac{1}{\delta-\chi}\right)\frac{W^{2}}{4^{i-1}}+W^{2}T^{\frac{\chi}{2}}\notag\\
&\quad+\frac{J^{2}}{4}T^{1-\frac{\chi}{2}}+2^{i-1}(F+JW)\frac{2}{1-2\chi}T^{\frac{1-\chi}{2}}\notag
\end{align}

From the convexity of $\|[g_t(x_t)]_+\|$, and combining~\eqref{x_update},~\eqref{final_decision},~\eqref{parameter}, and~\eqref{proof_72}, we have
\begin{align}\label{proof_73}
&\sum_{t=1}^{T}\|[h_t(x_t)]_+\| = \sum_{t=1}^{T}\|[h_t(\sum_{i=1}^{N}\rho_{i\text{,}t}x_{i\text{,}t})]_+\|\\
&\leq \sum_{t=1}^{T}\sum_{i=1}^{N}\rho_{i\text{,}t}\|[h_t(x_{i\text{,}t})]_+\|\leq \sum_{t=1}^{T}\sum_{i=1}^{N}\rho_{i\text{,}t}\|[h_t(x_{i\text{,}t})]_+\|_1\\
&\leq \sum_{i=1}^{N}\sum_{t=1}^{T}\|[h_t(x_{i\text{,}t})]_+\|_1\\
&\leq\sum_{i=1}^{N}\Bigg[\frac{J^2}{2}+\frac{3F}{2^{i-1}}+(1+\frac{1}{\delta-\chi})\frac{W^2}{4^{i-1}}+\frac{J^2}{4}T^{1-\frac{\chi}{2}}\\
&\quad+2^{i-1}(F+JW)\frac{2}{1-2\chi}T^{\frac{1}{2}-\frac{\chi}{2}}+W^2T^{\frac{\chi}{2}}\Bigg]\\
&\leq\left(\frac{J^2}{2}+W^2T^{\frac{\chi}{2}}+\frac{J^2}{4}T^{1-\frac{\chi}{2}}\right)\left(\frac{1}{2}\log_2(1+T)+2\right)+6F\\
&\quad+\frac{4}{3}\left(1+\frac{1}{\delta-\chi}\right)W^2+(F+JW)\frac{8}{1-2\chi}(T+1)^{\frac{1}{2}-\frac{\chi}{2}}\\
&=O(\log_2(T)T^{1-\frac{\chi}{2}})
\end{align}
Hence, we have finished the proof.

\bibliographystyle{IEEEtran}
\bibliography{IEEEabrv.bib,reference.bib}

\begin{IEEEbiography}[{\includegraphics[width=1in,height=1.25in,clip,keepaspectratio]{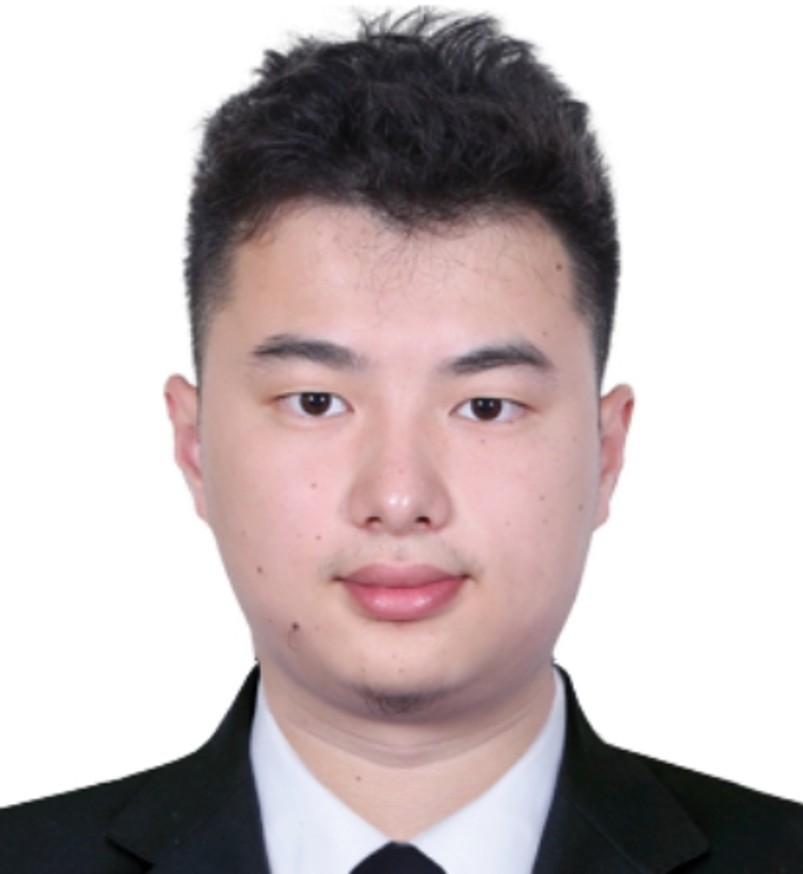}}]{Kaidi Huang}
    (Graduate Student Member, IEEE) was born in 2001. He received the B.S. degree from the Electrical Engineering Department, Tsinghua University, Beijing, China, in 2023, where he is currently pursuing the Ph.D. degree. His research interests include microgrid operation and control, optimization under uncertainty, and peer-to-peer energy trading for multi-energy systems.
\end{IEEEbiography}

\begin{IEEEbiography}[{\includegraphics[width=1in,height=1.25in,clip,keepaspectratio]{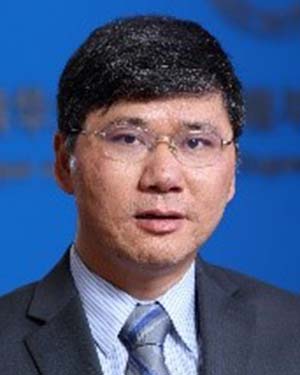}}]{Lin Cheng}
    (Senior Member, IEEE), was born in 1973. 
    He received the B.S. degree in electrical engineering from Tianjin University, Tianjin, China, in 1996 and the Ph.D. degree from Tsinghua University, Beijing, China, in 2001. 
    Currently, he is a tenured Professor in the Department of Electrical Engineering at Tsinghua University, serving as the deputy director of the State Key Laboratory of Power System Operation and Control, and is also a Fellow of IET.
    His research interests include operational reliability evaluation and application of power systems, 
    operation optimization of distribution systems with flexible resources, and perception and control of uncertainty in wide-area measurement systems.
\end{IEEEbiography}

\begin{IEEEbiography}[{\includegraphics[width=1in,height=1.25in,clip,keepaspectratio]{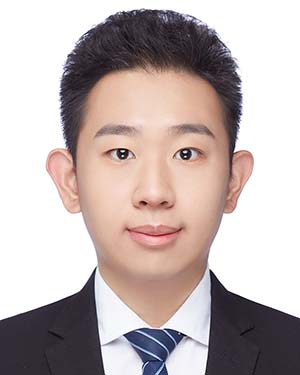}}]{Ning Qi}
    (S’16–M’23) was born in 1996. He received a B.S. degree in Electrical Engineering from Tianjin University, Tianjin, China, in 2018 and the Ph.D. degree in Electrical Engineering from Tsinghua University, Beijing, China, in 2023. He is currently a postdoctoral research scientist in Earth and Environmental Engineering at Columbia University. Before joining Columbia University, he was a visiting scholar at the Technical University of Denmark in 2022. He was a research associate in Electrical Engineering at Tsinghua University in 2024.  He received the Best Paper Award for \textit{IEEE PES General Meeting 2024} and the 2024 Outstanding Reviewer Award for \textit{IEEE Transactions on Smart Grid}. He is currently serving as the Associate Editor of \textit{IEEE Data Descriptions}, Youth Editorial Board Member for \textit{Power System Protection and Control}, \textit{Protection and Control of Modern Power Systems}, and \textit{Energy Conversion and Economics}. His current research focuses on flexibility modeling, optimization under uncertainty, and market design for power systems with generalized energy storage.
\end{IEEEbiography}

\begin{IEEEbiography}[{\includegraphics[width=1in,height=1.25in,clip,keepaspectratio]{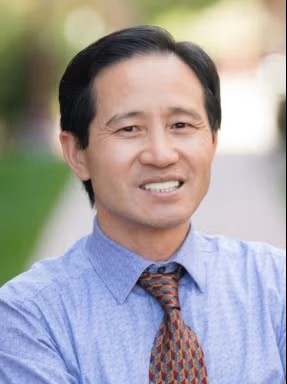}}]{David Wenzhong Gao} (S’00–M’02–SM’03–F’21) received his M.S. and Ph.D. degrees in electrical and computer engineering, specializing in electric power engineering, from Georgia Institute of Technology, Atlanta, USA, in 1999 and 2002, respectively. He is with the Department of Electrical and Computer Engineering, University of Denver, Colorado, USA. His teaching and research interests include renewable energy and distributed generation, microgrid, smart grid, power system protection, power electronics applications in power systems, power system modeling and simulation, and hybrid electric propulsion systems. He is an Associate Editor for \textit{IEEE Transactions on Intelligent Systems} and for \textit{Journal of Modern Power Systems and Clean Energy}. He was an editor of \textit{IEEE Transactions on Sustainable Energy}. He was the General Chair for the 48th North American Power Symposium (NAPS 2016) and the IEEE Symposium on Power Electronics and Machines in Wind Applications (PEMWA 2012).
\end{IEEEbiography}

\begin{IEEEbiography}[{\includegraphics[width=1in,height=1.25in,clip,keepaspectratio]{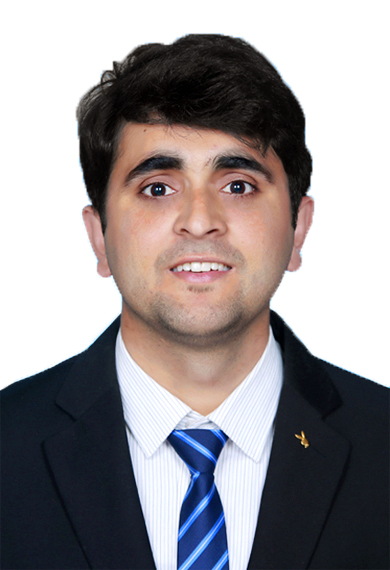}}]{Asad Mujeeb}
    (Graduate Student Member, IEEE) received the B.S. degree in Electrical Engineering from the Balochistan University of Information Technology, Engineering and Management Sciences, Quetta, Pakistan, in 2014, and the M.S. degree in Electrical Engineering from North China Electric Power University, Beijing, China, in 2020. He is currently working toward the Ph.D. degree in Electrical Engineering with Tsinghua University, Beijing, China. His research interests include Optimal Operation, and Scheduling of Virtual Power Plant (VPP), bidding strategies in electricity market, and power system uncertainties.
\end{IEEEbiography}

\begin{IEEEbiography}[{\includegraphics[width=1in,height=1.25in,clip,keepaspectratio]{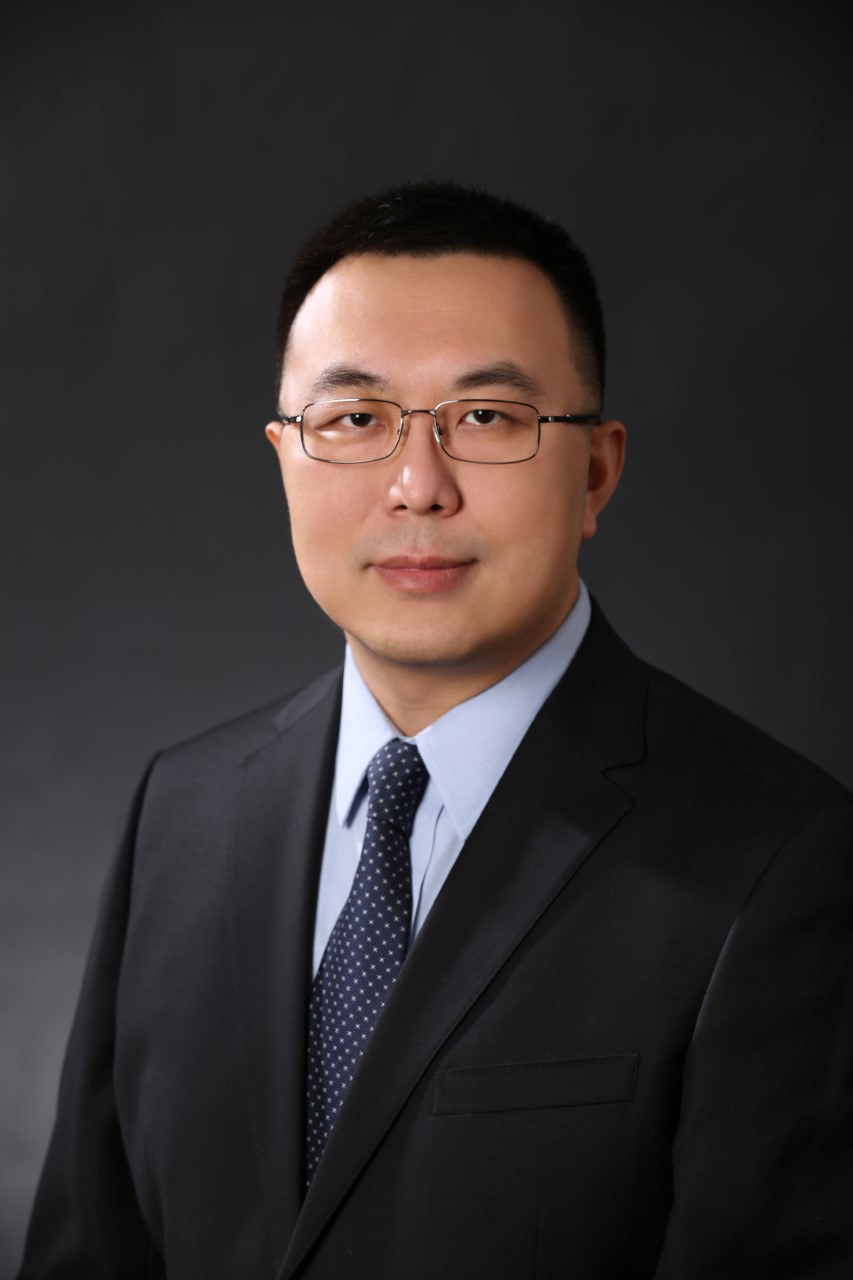}}]{Qinglai Guo}
    (SM'14-F'23) was born in Jilin City, Jilin Province in China on Mar. 6, 1979. He graduated from the Department of Electrical Engineering, Tsinghua University, Beijing, China, in 2000 with B.S degree. He received his PhD degree from Tsinghua University in 2005 where he is now a professor. He is now an IEEE Fellow, IET Fellow, and a CIGRE member, and is involved in 5 working groups of these organizations. Now he is TCPC of Energy Internet Coordinating Committee of IEEE PES, the co-chair of IEEE PES Work Group on "Energy Internet", IEEE PES Task Force on "Cyber-Physical Interdependence for Power System Operation and Control", and IEEE PES Task Force on "Voltage Control for Smart Grid". He is an editorial member of \textit{IEEE Transactions on Power Systems}, \textit{Renewable \& Sustainable Energy Reviews}, \textit{IEEE Transactions on Smart Grid}. His special fields of interest include smart grids, cyber-physical systems and electrical power control center applications.
\end{IEEEbiography}

\end{document}